\documentclass[twocolumn, tighten, numberedappendix,twocolappendix]{aastex631}
\usepackage{mathrsfs}
\usepackage{savesym}
\savesymbol{tablenum}
\usepackage[separate-uncertainty=true, multi-part-units=single]{siunitx}
\DeclareSIUnit\jansky{Jy}
\DeclareSIUnit\year{yr}
\restoresymbol{SIX}{tablenum}
\usepackage{tikz}
\usetikzlibrary{shapes,arrows}
\usepackage{verbatim}
\usepackage{cancel}
\usepackage{appendix}
\usepackage{amsmath}
\usepackage{comment}
\usepackage[normalem]{ulem}
\usepackage{lipsum}
\usepackage[utf8x]{inputenc}
\usepackage[encapsulated]{CJK}
\usepackage{ucs}
\newcommand{\cntext}[1]{\begin{CJK}{UTF8}{gbsn}#1\end{CJK}}
\usepackage{newtxtext}
\usepackage[varg]{newtxmath}
\usepackage{ulem}
\shortauthors{Appel et al.}
\usepackage{devanagari}
\usepackage{tablefootnote}

\begin{document}

\title{Calibration of Transition-edge Sensor (TES) bolometer arrays with application to CLASS}


\author[0000-0002-8412-630X]{John W.~Appel}
\affiliation{Department of Physics and Astronomy, Johns Hopkins University, 
3701 San Martin Drive, Baltimore, MD 21218, USA}

\correspondingauthor{John W.~Appel}
\email{jappel3@jhu.edu}

\author[0000-0001-8839-7206]{Charles L.~Bennett}
\affiliation{Department of Physics and Astronomy, Johns Hopkins University, 
3701 San Martin Drive, Baltimore, MD 21218, USA}

\author{Michael K. Brewer}
\affiliation{Department of Physics and Astronomy, Johns Hopkins University,
3701 San Martin Drive, Baltimore, MD 21218, USA}

\author[0000-0001-8468-9391]{Ricardo Bustos}
\affiliation{Facultad de Ingenier\'{i}a, Universidad Cat\'{o}lica de la Sant\'{i}sima Concepci\'{o}n, Alonso de Ribera 2850, Concepci\'{o}n, Chile}

\author{Manwei Chan}
\affiliation{Department of Physics and Astronomy, Johns Hopkins University, 3701 San Martin Drive, Baltimore, MD 21218, USA}

\author[0000-0003-0016-0533]{David T. Chuss}
\affiliation{Department of Physics, Villanova University, 800 Lancaster Avenue, Villanova, PA 19085, USA}

\author[0000-0002-7271-0525]{Joseph Cleary}
\affiliation{Department of Physics and Astronomy, Johns Hopkins University, 3701 San Martin Drive, Baltimore, MD 21218, USA}

\author[0000-0002-0552-3754]{Jullianna D. Couto}
\affiliation{Department of Physics and Astronomy, Johns Hopkins University, 3701 San Martin Drive, Baltimore, MD 21218, USA}

\author[0000-0002-1708-5464]{Sumit Dahal}
\affiliation{NASA Goddard Space Flight Center, 8800 Greenbelt Road, Greenbelt, MD 20771, USA}
\affiliation{Department of Physics and Astronomy, Johns Hopkins University, 3701 San Martin Drive, Baltimore, MD 21218, USA}

\author[0000-0003-3853-8757]{Rahul Datta}
\affiliation{Department of Physics and Astronomy, Johns Hopkins University, 3701 San Martin Drive, Baltimore, MD 21218, USA}

\author[0000-0002-3592-5703]{Kevin Denis}
\affiliation{NASA Goddard Space Flight Center, 8800 Greenbelt Road, Greenbelt, MD 20771, USA}

\author[0000-0001-6976-180X]{Joseph Eimer}
\affiliation{Department of Physics and Astronomy, Johns Hopkins University, 3701 San Martin Drive, Baltimore, MD 21218, USA}

\author[0000-0002-4782-3851]{Thomas Essinger-Hileman}
\affiliation{NASA Goddard Space Flight Center, 8800 Greenbelt Road, Greenbelt, MD 20771, USA}
\affiliation{Department of Physics and Astronomy, Johns Hopkins University, 3701 San Martin Drive, Baltimore, MD 21218, USA}

\author[0000-0003-1248-9563]{Kathleen Harrington}
\affiliation{Department of Astronomy and Astrophysics, University of Chicago, 5640 South Ellis Avenue, Chicago, IL 60637, USA}
\affiliation{Department of Physics and Astronomy, Johns Hopkins University, 3701 San Martin Drive, Baltimore, MD 21218, USA}

\author[0000-0001-7466-0317]{Jeffrey Iuliano}
\affiliation{Department of Physics and Astronomy, University of Pennsylvania, 209 South 33rd Street, Philadelphia, PA 19104, USA}

\author[0000-0002-4820-1122]{Yunyang Li}
\affiliation{Department of Physics and Astronomy, Johns Hopkins University, 3701 San Martin Drive, Baltimore, MD
21218, USA}

\author[0000-0003-4496-6520]{Tobias~A. Marriage}
\affiliation{Department of Physics and Astronomy, Johns Hopkins University, 3701 San Martin Drive, Baltimore, MD
21218, USA}

\author[0000-0002-5247-2523]{Carolina N\'{u}\~{n}ez}
\affiliation{Department of Physics and Astronomy, Johns Hopkins University, 3701 San Martin Drive, Baltimore, MD 21218, USA}

\author[0000-0003-2838-1880]{Keisuke Osumi}
\affiliation{Department of Physics and Astronomy, Johns Hopkins University, 3701 San Martin Drive, Baltimore, MD 21218, USA}

\author[0000-0002-0024-2662]{Ivan L.~Padilla}
\affiliation{Department of Physics and Astronomy, Johns Hopkins University,
3701 San Martin Drive, Baltimore, MD 21218, USA}

\author[0000-0002-4436-4215]{Matthew~A. Petroff}
\affiliation{Department of Physics and Astronomy, Johns Hopkins University, 3701 San Martin Drive, Baltimore, MD 21218, USA}
\affiliation{Center for Astrophysics, Harvard \& Smithsonian, 60 Garden Street, Cambridge, MA 02138, USA}

\author[0000-0003-4189-0700]{Karwan Rostem}
\affiliation{NASA Goddard Space Flight Center, 8800 Greenbelt Road, Greenbelt, MD 20771, USA}

\author[0000-0003-3487-2811]{Deniz A. N. Valle}
\affiliation{Department of Physics and Astronomy, Johns Hopkins University, 3701 San Martin Drive, Baltimore, MD 21218, USA}

\author[0000-0002-5437-6121]{Duncan J. Watts}
\affiliation{Institute of Theoretical Astrophysics, University of Oslo, P.O. Box 1029 Blindern, N-0315 Oslo, Norway}
\affiliation{Department of Physics and Astronomy, Johns Hopkins University, 3701 San Martin Drive, Baltimore, MD 21218, USA}

\author[0000-0003-3017-3474]{Janet L. Weiland}
\affiliation{Department of Physics and Astronomy, Johns Hopkins University, 3701 San Martin Drive, Baltimore, MD 21218, USA}

\author[0000-0002-7567-4451]{Edward J. Wollack}
\affiliation{NASA Goddard Space Flight Center, 8800 Greenbelt Road, Greenbelt, MD 20771, USA}

\author[0000-0001-5112-2567]{Zhilei Xu (\cntext{徐智磊}$\!\!$)}
\affiliation{ MIT Kavli Institute, Massachusetts Institute of Technology, 77 Massachusetts Avenue, Cambridge, MA 02139, USA}
\affiliation{Department of Physics and Astronomy, Johns Hopkins University, 3701 San Martin Drive, Baltimore, MD 21218, USA}

\published{October 7, 2022}
\submitjournal{\apjs}

\begin{abstract}

The current and future cosmic microwave background (CMB) experiments fielding kilo-pixel arrays of transition-edge sensor (TES) bolometers require accurate and robust gain calibration methods.
We simplify and refactor the standard TES model to directly relate the detector responsivity calibration and optical time constant to the measured TES current $I$ and the applied bias current $I_{\mathrm{b}}$.
The calibration method developed for the Cosmology Large Angular Scale Surveyor (CLASS) TES bolometer arrays relies on current versus voltage ($I$-$V$) measurements acquired daily prior to CMB observations. 
By binning Q-band (\SI{40}{\giga\hertz}) $I$-$V$ measurements by optical loading, we find that the gain calibration median standard error within a bin is 0.3\%.
We test the accuracy of this ``$I$-$V$ bin''  detector calibration method by using the Moon as a photometric standard. The ratio of measured Moon amplitudes between the detector pairs sharing the same feedhorn indicates a TES calibration error of 0.5\%. 
We also find that, for the CLASS Q-band TES array, calibrating the response of individual detectors based solely on the applied TES bias current accurately corrects TES gain variations across time but introduces a bias in the TES calibration from data counts to power units. Since the TES current bias value is set and recorded before every observation, this calibration method can always be applied to the raw TES data and is not subject to $I$-$V$ data quality or processing errors.

\keywords{\href{http://astrothesaurus.org/uat/322}{Cosmic microwave background radiation (322)}; \href{http://astrothesaurus.org/uat/435}{Early Universe (435)}; \href{http://astrothesaurus.org/uat/1146}{Observational Cosmology (1146)}; \href{http://astrothesaurus.org/uat/799}{Astronomical instrumentation (799)}; \href{http://astrothesaurus.org/uat/1127}{Polarimeters (1127)}; \href{http://astrothesaurus.org/uat/259}{CMBR Detectors (259)}}

\end{abstract}

\section{Introduction}\label{sec:intro}

Cosmic microwave background (CMB) observatories \citep{spt3g, bicep3, spider_2018, class_kh, pb2, piper, advact} are mapping the sky with kilo-pixel arrays of transition-edge sensor (TES) detectors, and telescopes fielding even larger TES arrays will begin observing this decade \citep{so, cmbs4, litebird}.
Calibrating detector time-ordered data (TOD) to a standard unit is often the first step in processing astrophysical data sets. The calibration method must be accurate, to suppress systematic errors in the final results, and robust, to be applicable to a vast majority of observations. Data that cannot be calibrated are excluded, reducing the sensitivity of the instrument.

 Current CMB telescopes operating TES bolometers have corrected for individual detector steady-state gain (i.e., DC gain) fluctuations using four methods based on: elevation nods~\citep{bicep_calib}, chopped
thermal sources~\citep{polarbear_calib, spt3g_calib}, $I$-$V$ data (i.e., load curves)~\citep{act_rolo, abs, bicep_calib}, and electrical bias step~\citep{niemack_thesis, rahlin_thesis,bias_steps_spider}. 
The first two methods calibrate the TES gain to the brightness of an external source. Elevation nods measure the change in atmospheric emission with elevation, while chopped thermal sources are mounted in the telescope's optical path and turned on/off on short time scales.
These methods directly measure the optical response of the detectors but change the nominal observing state of the telescopes and thus use a portion of the observing time.
Load curves measure the TES response to sweeping the bias voltage over a wide range, driving the TES from the normal state, to operating on transition, to superconducting. The bias step method drives small changes in TES bias current on top of the operating bias.
Many experiments acquire $I$-$V$ data before an observing run to choose the optimal operating bias for the detectors, with a run time of about one minute. Therefore, the $I$-$V$ gain calibration is obtained from data already collected by the standard operation of the TES detectors.
Bias steps may take only a few seconds and can be acquired during standard observations. 
These last two methods do not directly measure the optical response of the detectors, but infer it from the measured TES response to changes in bias current through our understanding of the TES bolometer electrothermal model. 
The novel $I$-$V$~bin calibration method presented here combines daily load curve data across multiple observing years to improve the gain calibration for each day-long observing run.
The improvement in gain calibration is achieved without decreasing the CMB observing efficiency.
Future CMB space missions deploying TES bolometers~\citep{litebird,pico} will likely rely on the Earth-velocity modulation of the CMB dipole as a gain calibrator~\citep{wmap_bennet_calib,planck_calib}; nevertheless, their TES bolometer arrays may require periodic re-biasing based on load curves, providing an independent gain calibration method.
The unique characteristics of any CMB telescope deploying TES bolometers combined with its observing strategy will determine which TES calibration methods are necessary to achieve the desired accuracy and control over systematic effects.

By mapping the polarization of the CMB over angular scales between 1$^\circ \lesssim \theta \leq$ 90$^\circ$, CLASS aims to test the paradigm of inflation ~\citep{guth:1981, sato:1981,linde:1982,starobinsky:1982,albrecht/steinhardt:1982, planck_2018_inflation} and probe the epoch of cosmic reionization~~\citep{Hinshaw2013,planck_2018_cosmo_param}.
The CLASS polarimeters measure the microwave sky in bands centered at approximately \SIlist{40;90;150;220}{\giga\hertz} from an altitude of \SI{5200}{\meter} above sea level in the Atacama desert of northern Chile~\citep{tom_spie,class_kh} inside the Parque Astron\'omico Atacama~\citep{parque_atacama}. 
The Q-band (\SI{40}{\giga\hertz}) telescope~\citep{joseph_SPIE,spie_jappel} is sensitive to Galactic synchrotron emission, whereas the G-band (dichroic \SIlist{150;220}{\giga\hertz}) telescope~\citep{sumit_hf} is sensitive to Galactic dust emission.
Two W-band (\SI{90}{\giga\hertz}) telescopes~\citep{sumit_w} provide sensitivity to the CMB polarized signal near the Galactic foreground minimum $\sim$ 70~GHz \citep{bennett_2013,planck_diffuse}. 
The CLASS telescopes employ variable-delay polarization modulators (VPM) as their first optical element to increase stability and mitigate instrumental polarization~\citep{Miller2015,chuss_vpm_2012,katie_Qvpm_2021}.
A co-moving ground shield limits terrestrial signal contamination.
Telescope boresight rotation provides an additional level of sky signal polarization modulation by the instrument. 
The CLASS survey is forecast to constrain the optical depth to reionization to near the cosmic variance limit and the inflationary tensor-to-scalar ratio to $r\approx 0.01$~\citep{Watts2015,Watts2018}.
The optical depth is the least constrained $\Lambda$CDM parameter, and new constraints from measurements at spherical-harmonic multipole moments $\ell \lesssim 20$ are important for realizing the full potential of cosmological probes of neutrino masses \citep{Allison2015,Watts2018}.

This article describes the detector calibration method applied to the CLASS data. 
Section~\ref{sec:model} presents the TES bolometer model used to interpret $I$-$V$ data and extract detector responsivity calibration factors. 
Section~\ref{sec:iv_analysis_ivbin} introduces a novel calibration method that leverages the many $I$-$V$ data sets acquired over one or multiple observing years to optimize the calibration of each CMB scan set corresponding to a single $I$-$V$ data set (each CLASS CMB scan set is approximately one day long).
Section~\ref{sec:class_calib} discusses the detector calibration pipeline applied to CLASS data, from raw detector data to sky temperature units. It includes single detector calibration, relative calibration between detectors, gain corrections due to atmospheric opacity, and absolute calibration to CMB thermodynamic units. It also summarizes the various time scales of the experiment including: data rate, detector time constants, and calibration periods.
Finally, Section~\ref{sec:calib_test} tests the validity and accuracy of the TES calibration methods considered in this paper by using observations of the Moon and the detector noise model as calibration references.

\section{TES bolometer model}\label{sec:model}

\begin{figure}
    \includegraphics[width=1.0\linewidth]{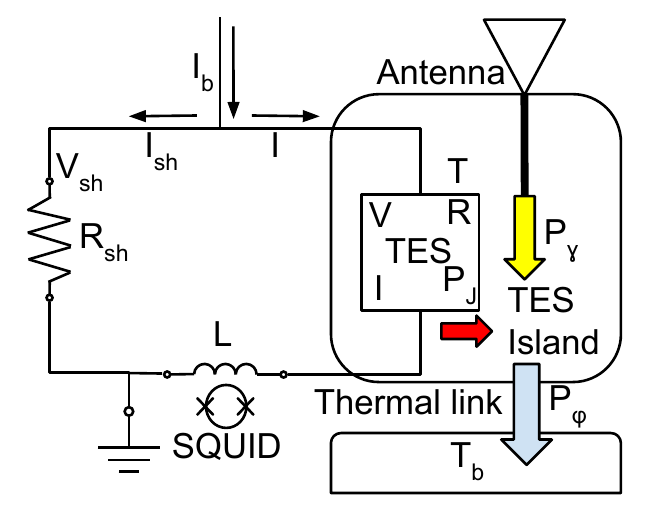}
    \caption{Electrothermal diagram of a TES bolometer.
    A constant bias current ($I_{\mathrm{b}}$) flows through the TES and 
    a shunt resistor ($R_{\mathrm{sh}}\ll R$). The circuit keeps the TES voltage ($V$) nearly constant; therefore, a change in TES resistance ($R$) leads to a change in TES current ($I$). Changes in $I$ are measured with a superconducting quantum interference device (SQUID) connected in series with the TES.
    Optical power ($P_{\gamma}$), coupled to the TES bolometer through the antenna, heats the TES island, causing a change in $R$ and hence a current signal measured by the SQUID. 
    At equilibrium, the TES operates on its superconducting transition  at temperature $T_{\mathrm{c}}$, and the sum of the bias power ($P_{\mathrm{J}}$) and $P_{\gamma}$ dissipated on the island equals the power ($P_{\upvarphi}$) transported to the bath (at temperature $T_{\mathrm{b}}$) by phonons.
    The interaction between the electrical circuit biasing the TES and the thermal circuit controlling the power flow across the TES island to the bath forms the electrothermal model that governs the TES detector behavior. The coupling between these two systems is determined by the design and intrinsic properties of TES superconducting film, as well as the TES island geometry and its material properties.  
    }
\label{fig:thermo_elec}
\end{figure}

In this section, we develop a TES bolometer model (see Figure~\ref{fig:thermo_elec}) based on the following assumptions:
\begin{enumerate}
\item The TES operates at equilibrium with the power dissipated on the TES island matching the power transported by phonons from the island to the cold bath.
\item The response of the TES electrical circuit is instantaneous compared to the TES thermal circuit (i.e., the low electrical-inductance limit).
\end{enumerate}
Our aim is to model the TES with a negative feedback circuit (see Figure~\ref{fig:tes_bol}) and use it to extract TES parameters from $I$-$V$ data. 
This approach circumvents solving the coupled differential equations describing the TES electrothermal circuit.
Instead, we directly map the slope of $I$-$V$ data to the TES zero-frequency responsivity.

\subsection{Bolometer power}

A TES bolometer can be modeled as a coupled electrothermal system \citep{irwin_hilton,mather_bolometer,richards_bolometer}, as depicted in Figure~\ref{fig:thermo_elec}. 
At equilibrium, the sum of the TES bias power ($P_{\mathrm{J}}$) and optical signal ($P_{\gamma}$) dissipated on the TES island equals the phonon power ($P_{\upvarphi}$) that flows from the ``hot'' island to the ``cold'' bath through the silicon beam connecting them. 
This flow of power can be in the form of ballistic or diffuse phonons. The architecture and material properties of the CLASS bolometers~\citep{karwan_legs,goddard_det} imply ballistic phonons and hence power $P_{\upvarphi}$ given by:
\begin{equation}
P_{\upvarphi} = \kappa (T^n-T_\mathrm{b}^n) \approx \kappa(T_{\mathrm{c}}^n-T_\mathrm{b}^n), 
\label{eqn_Pkappa}
\end{equation}
where $n=4$, and $\kappa$ is the thermal conductance constant set by the composition, geometry, and fabrication process of the beam~\citep{karwan_legs}.
$T_{\mathrm{b}}$ is the bath temperature, while $T$ is the temperature of the device, which is restricted to a narrow range around the fiducial critical temperature $T_\mathrm{c}$, below which the resistance drops abruptly to zero.
On the superconducting transition, the temperature $T(I,R)$ is a function of TES resistance $R$ and TES current $I$.\footnote{Unlike the more commonly used function $R(I,T)$, the function $T(I,R)$ is single valued only on the superconducting transition. This is not limiting as all considerations herein concern functions on transition.}

The thermal conductance ($G$) between the bolometer island and surrounding silicon frame is defined as
\begin{equation}
G = \left.\frac{\partial P_{\upvarphi}}{\partial T}\right|_{T_\mathrm{b}} = n \kappa T^{n-1}
\approx n \kappa T_{\mathrm{c}}^{n-1}
.
\label{eqn:g}
\end{equation}

\begin{figure*}
\begin{center}
    \includegraphics[width=1.0\linewidth, trim = {0cm, 0cm, 0cm, 0cm}]{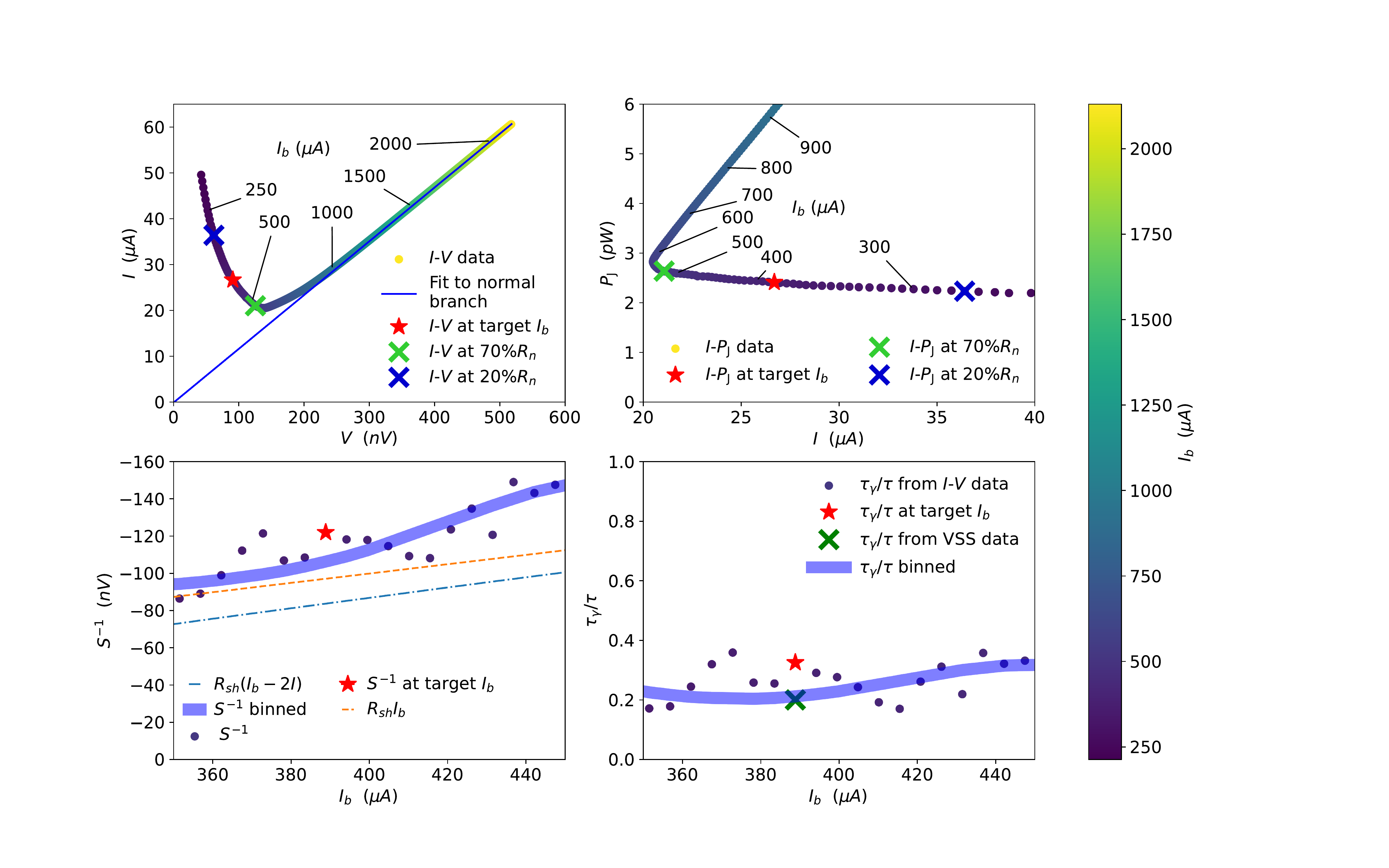}
\end{center}
    \caption{Example $I$-$V$ data from a CLASS Q-band TES. \textbf{Top left}: 
    Plot of TES voltage ($V$) versus TES current ($I$) response to sweeping the TES bias current ($I_{\mathrm{b}}$) from high to low.
    The blue line fits the section of the $I$-$V$ where the TES operates as an Ohmic/normal resistor. The slope of the line fit is equal to the inverse of the TES normal resistance ($R_{\mathrm{n}}$), and the y-intercept offset is subtracted from the raw data to calibrate the $I$-$V$ to absolute current units.
    The red star marks $I$ and $V$ at the target $I_{\mathrm{b}}$ set for CMB observation.
    The green cross marks $I$ and $V$ at 70\% of $R_{\mathrm{n}}$, and the blue cross marks them at 20\% of $R_{\mathrm{n}}$. 
    \textbf{Top right}: 
    Plot of $I$-$V$ data expressed in terms of bias power ($P_{\mathrm{J}}$) on the y-axis, and $I$ on the x-axis.
    The symbols are the same as in the top-left plot. 
    For an ideal TES, $P_{\mathrm{J}}$ would be constant across the TES transition and hence at all $I$. For the CLASS detectors, $P_{\upvarphi}$ decreases across the TES transition with increasing $I$.
    \textbf{Bottom left}: The blue points mark the derived responsivity from the $I$-$V$ data shown in the top-left plot (see Section~\ref{sec:iv_resp}). The dashed blue line is the ideal TES responsivity ($S^*$) approximation described in Section~\ref{sec:ideal_resp}, while the dashed orange line is the $S_{I_{\mathrm{b}}}$ responsivity approximation based on $I_{\mathrm{b}}$. The red star marks the responsivity at the target $I_{\mathrm{b}}$ value set for the observations acquired after this particular $I$-$V$ was measured. The shaded blue region is centered on the mean $I$-$V$~bin responsivity (see Section~\ref{sec:iv_analysis_ivbin}), and the contour width is the  1-$\sigma$ distribution of $I$-$V$ responsivity values in the corresponding $P_{\mathrm{sat}}$ bin. Typically for a CLASS TES, the $I$-$V$ responsivity is closer to the $S_{I_{\mathrm{b}}}$ responsivity estimate than to the $S^*$ ideal TES responsivity estimate. This scenario can reverse for TES with different characteristics. \textbf{Bottom right}: The blue points mark the estimated $\tau_{\gamma}/\tau$ from the $I$-$V$ data shown in the top-left plot (see Section~\ref{sec:iv_analysis_tau}). The green cross plots the measured $\tau_{\gamma}/\tau$ ratio based on the time constant extracted from the VPM synchronous signal (VSS)~\citep{katie_thesis}.}
\label{fig:iv}
\end{figure*}

An $I$-$V$ measurement consists of applying a high bias current ($I_{\mathrm{b}}$) across the bolometer, heating the TES above its superconducting transition, and then decreasing the bias slowly (about 60 seconds from maximum to zero $I_{\mathrm{b}}$). 
A typical $I$-$V$ curve is shown in Figure~\ref{fig:iv}. 
Initially, the TES current ($I$) decreases linearly with $I_{\mathrm{b}}$ as expected for a normal resistor with resistance $R_{\mathrm{n}}$ (i.e., the normal branch).
At low enough $I_{\mathrm{b}}$, the TES island temperature decreases to $T_{\mathrm{c}}$, and the TES begins to function on its superconducting transition, where $I$ increases as $I_{\mathrm{b}}$ decreases. 
On transition, the TES electrothermal feedback increases $I$ to maintain the equilibrium Joule power dissipation on the TES island as the voltage bias decreases with decreasing bias current.
As $I_{\mathrm{b}}$ approaches zero current, the equilibrium island temperature falls below $T_{\mathrm{c}}$, making the TES a superconducting short (i.e., the superconducting branch).

It is important to note that the SQUID readout measures changes in TES current ($\Delta I$) and not absolute TES current ($I$).
$I$-$V$ curves are calibrated to absolute TES current units by linearly extrapolating the normal and/or superconducting branches and subtracting an offset to make these extrapolations go through the origin.
Linearly extrapolating the normal branch section of the $I$-$V$ that is far from the superconducting transition reduces the uncertainty on the offset. For CLASS $I$-$V$s, the median per-detector offset  uncertainty is 0.4\% (1-$\sigma$). Combining $I$-$V$ data acquired under similar observing conditions further reduces the uncertainty on the absolute TES current calibration (see Section~\ref{sec:iv_analysis_ivbin}).

The TES resistance is connected in parallel with a shunt resistor ($R_{\mathrm{sh}}$); hence, the TES DC voltage $V$ equals the shunt voltage $V_{\mathrm{sh}}$, and the TES circuit splits $I_{\mathrm{b}}$ into $I$ and $I_{\mathrm{sh}}$ (see circuit diagram in Figure~\ref{fig:thermo_elec}).
The electrical bias power dissipated on the TES ($P_{\mathrm{J}}$) is obtained from:
\begin{equation}
P_{\mathrm{J}} \equiv V I = V_{\mathrm{sh}} I = R_{\mathrm{sh}} I_{\mathrm{sh}} I = R_{\mathrm{sh}} (I_{\mathrm{b}}-I) I.
\label{p_r_tes}
\end{equation}
The bias current $I_{\mathrm{b}}$ is set by the user, and the TES current $I$ is measured from the calibrated $I$-$V$.
The resistance of the shunt is measured by fitting a Johnson noise model to the noise spectra of the TES channel when the TES is in the superconducting state (i.e., zero bias current and $T_{\mathrm{b}} < T_{\mathrm{c}}$). We find that the measured shunt resistances for CLASS match the fabrication targets (see Table~\ref{tab:class_mce}). 
The TES voltage bias circuit relates $R$ to $I_{\mathrm{b}}$ and $I$ through
\begin{equation}
    R = \frac{V}{I}= \frac{V_{\mathrm{sh}}}{I} =  \left(\frac{I_{\mathrm{b}}}{I}-1 \right) R_{\mathrm{sh}}.
    \label{eqn:R_to_Ib}
\end{equation}
This relationship allows us to express $T(I,R)$ as a function of the measured variables $T(I, I_\mathrm{b})$. It follows that $P_{\upvarphi}$ can also be written in terms of $I$ and $I_{\mathrm{b}}$.

At equilibrium, the phonon power outflow equals the sum of the input electrical bias power $P_{\mathrm{J}}$ and the optical power $P_{\gamma}$ dissipated on the TES bolometer
\begin{equation}
P_{\upvarphi} = P_{\gamma}+P_{\mathrm{J}}.
\label{p_eq2}
\end{equation}
$P_{\gamma}$ is a combination of optical power coupled through the detector antennas in-band, out-of-band power that leaks through the band-defining on-chip filters, and stray light that reaches into the TES cavity and couples directly to the TES island.
The CLASS detectors were designed to strongly suppress coupling to stray light and out-of-band optical power~\citep{goddard_det,goddard_crowe}.

Before deploying each telescope, we conduct in-lab dark tests with the detector focal plane fully enclosed in a dark \SI{1}{\kelvin} cavity (typically operating at \SI{0.65}{\kelvin}). 
Bolometer data acquired during dark tests satisfy the condition: $P_{\gamma} \approx 0$, so that $ P_{\mathrm{J}}\approx P_{\upvarphi}$. The dark cavity radiates $\sim$\SI{0.02}{\pico\watt} of optical power in the CLASS \SI{40}{\giga\hertz} band and lower amounts at the higher-frequency CLASS bands.
The dark cavity optical loading is negligible compared to the bias power ($\sim$\SI{10}{\pico\watt}) of the CLASS detectors~\citep{sumit_4year_det}.
$P_{\gamma}$ is measured during observations by subtracting power $P_{\mathrm{J}}$ (extracted from on-sky $I$-$V$s) from $P_{\upvarphi}$ (deduced from dark $I$-$V$s), with both measured at the same $T_{\mathrm{b}}$\footnote{Changes to the magnetic field environment near the TES between the lab and site configurations could lead to variations in $T_{\mathrm{c}}$ that would introduce systematic errors to the $P_{\gamma}$ measurements. Multiple layers of high-permeability Amumetal 4K (\url{https://www.amuneal.com}) in the CLASS cryostats suppress the ambient magnetic field by a factor of $\sim$100~\citep{aamir_thesis}, improving the accuracy of $P_{\gamma}$ measurements from $I$-$V$s.}.
In lab with zero optical loading, Equation~\ref{eqn_Pkappa} can be used to fit for $\kappa$ and $T_{\mathrm{c}}$ by measuring $P_{\mathrm{J}}\approx P_{\upvarphi}$ at multiple $T_{\mathrm{b}}$ temperatures below $T_{\mathrm{c}}$.

In an ideal TES, the superconducting transition $R$ vs $T$ is sharp and independent of $I$. 
In this ideal case, the power $P_{\upvarphi}$ conducted to the bath from a TES biased on transition is constant. 
The superconducting temperature of the CLASS TES depends weakly on $I$ and $I_{\mathrm{b}}$; hence, $P_{\upvarphi}$ decreases by a small amount as the detector is biased lower on the transition (Figure~\ref{fig:iv} shows $P_{\mathrm{J}}$ as a function of $I$).
It is then useful to consider, for each $I$-$V$, the value of $P_{\mathrm{J}}$ measured high on the transition where  $R\approx R_\mathrm{n}$ and $I$ is near its minimum value.
In particular for CLASS, we choose to define the TES saturation power $P_{\mathrm{sat}}$ as $P_{\mathrm{J}}$ computed at $70\%~R_{\mathrm{n}}$.  
In dark tests ($P_{\mathrm{J}}\approx P_{\upvarphi}$), $P_{\mathrm{sat}}$ can be interpreted as the maximum optical power that the TES island can absorb while maintaining its temperature on transition at $T_{\mathrm{c}}$.

\subsection{Responsivity at the zero-frequency limit}
\label{sec:iv_analysis_responsivity}

During a single CMB observation schedule ($\sim$1 day), a CLASS detector operates at the same $I_{\mathrm{b}}$ and typically observes $P_{\gamma}$ fluctuations on the order of a few percent of $P_{\mathrm{J}}$, driven primarily by changes in the opacity and temperature of the atmosphere.
Therefore, we focus on calibrating small $\Delta I$ signals to the corresponding change in optical power dissipated on the TES island $\Delta P_{\gamma}$ at a set $I_{\mathrm{b}}$. 
This calibration factor is called responsivity ($S$) and varies across detectors and/or observing conditions. 
The responsivity equation derived in \citet{irwin_hilton} evaluated at the zero-frequency limit (i.e., DC responsivity) is
\begin{equation}
    S(0) = - \frac{1}{I R} \left(\frac{L}{\tau_{\mathrm{el}} R \mathscr{L}_I}+ \left(1-\frac{R_{\mathrm{sh}}}{R} \right)\right)^{-1}.
\label{resp_ih}
\end{equation}
It depends on TES electrical time constant ($\tau_{\mathrm{el}}$) and the TES loop gain ($\mathscr{L}_I$) that typically require complex impedance measurements to constrain~\citep{jappel_truce_complex}. $L$ is the TES loop inductance.
The TES loop gain $\mathscr{L}_I$ is defined as
\begin{equation}
    \mathscr{L}_I = \frac{P_{\mathrm{J}}\alpha}{G T} =  \frac{P_{\mathrm{J}}}{G T} \left.\frac{T}{R}\frac{\partial R}{\partial T}\right|_{I}
    = \frac{I^2}{G}\left.\frac{\partial R}{\partial T}\right|_{I},
    \label{eqn_loop_gain}
\end{equation}
and $\tau_{\mathrm{el}}$ as
\begin{equation}
    \tau_{\mathrm{el}} = \frac{L}{R_{\mathrm{sh}}+R(1+\beta)},
    \label{eqn_tel}
\end{equation}
where  $\alpha = T/R \left.\partial R/\partial T \right|_{I}$, and $\beta= I/R \left.\partial R/\partial I\right|_{T}$.

Equation~\ref{resp_ih} can be expressed as
\begin{equation}
\begin{split}
    S^{-1}(0) = R_{\mathrm{sh}} I \left(1-\frac{R}{R_{\mathrm{sh}}}-\frac{1}{\mathscr{L}_I}-\frac{R}{R_{\mathrm{sh}}}\frac{(1+\beta)}{\mathscr{L}_I}\right) \\ 
    = R_{\mathrm{sh}} I \left(1-\mathscr{L}_I^{-1}\right)  \left(1-\frac{R}{R_{\mathrm{sh}}} \left(\frac{1+\beta+\mathscr{L}_I}{\mathscr{L}_I-1} \right)\right).
\end{split}
\label{eqn:resp_ih2}
\end{equation}

Below, we present an alternate approach to compute the TES DC responsivity~\citep{jappel_thesis} that relies on the slope and absolute calibration of $I$-$V$ curves. 
In the following sections, references to ``responsivity'' refer to responsivity in the zero-frequency limit unless otherwise specified. 

\subsubsection{Ideal TES DC responsivity}
\label{sec:ideal_resp}

Responsivity is defined as the change in TES current $\Delta I$ for a small change in $\Delta P_\gamma$. We can expand the latter as
\begin{align}
    \Delta P_\gamma &= \Delta P_\upvarphi - \Delta P_J \\
    &= \left(\left.\frac{\partial P_\upvarphi}{\partial I}\right|_{I_{\mathrm{b}}}\Delta I +  \left.\frac{\partial P_\upvarphi}{\partial I_{\mathrm{b}}}\right|_{I}\Delta I_{\mathrm{b}}\right) \notag\\ 
    & \,\,\,\, - \left(\left.\frac{\partial P_J}{\partial I}\right|_{I_{\mathrm{b}}}\Delta I + \left.\frac{\partial P_J}{\partial I_{\mathrm{b}}}\right|_{I}\Delta I_{\mathrm{b}}\right) \notag
\end{align}
Under observing conditions, the bias current is constant ${\Delta I_{\mathrm{b}}=0}$. Therefore the inverse responsivity can be expressed as
\begin{equation}
    S^{-1} = \frac{\Delta P_\gamma}{\Delta I}  = \left.\frac{\partial P_{\upvarphi}}{\partial I}\right|_{I_{\mathrm{b}}}-\left.\frac{\partial P_{\mathrm{J}}}{\partial I}\right|_{I_{\mathrm{b}}}.
\label{resp_2}
\end{equation}


In the ideal case where $P_{\upvarphi}$ is approximately constant across the transition, the first term on the right hand side of Equation~\ref{resp_2} is zero, and using Equation~\ref{p_r_tes} to substitute for $P_{\mathrm{J}}$ leads to
\begin{equation}
\frac{1}{S^*} = \left.\frac{\partial P_{\gamma}}{\partial I}\right|_{I_{\mathrm{b}}, P_{\upvarphi}} =-\left.\frac{\partial P_{\mathrm{J}}}{\partial I}\right|_{I_{\mathrm{b}}}=- R_{\mathrm{sh}}(I_{\mathrm{b}}-2 I),
\label{resp_3}
\end{equation}
where $S^*$ denotes DC responsivity in the limit of an ideal TES where $P_{\upvarphi}$ is constant across the transition. 
For the more general case where $P_{\upvarphi}$ is not constant, we will show in Section~\ref{sec:iv_resp} that the relevant contributions to the first term of Equation~\ref{resp_2} can be estimated from the $I$-$V$ curve measurement.
\subsubsection{$I_{\mathrm{b}}$ DC Responsivity}
To apply a voltage bias across the TES, we require that $R_{\mathrm{sh}} \ll R$, and hence $I \ll I_{\mathrm{b}}$. Therefore, a good approximation to Equation~\ref{resp_3} is
\begin{equation}
S^{-1}_{I_{\mathrm{b}}}=- R_{\mathrm{sh}} I_{\mathrm{b}},
\label{resp_Ib}
\end{equation}
where $S_{I_{\mathrm{b}}}$ denotes the $I_{\mathrm{b}}$ DC responsivity approximation in the limit where $P_{\upvarphi}$ is constant across the transition and $R_{\mathrm{sh}} \ll R$.
Approximating responsivity through Equation~\ref{resp_Ib} has the crucial advantage that it only depends on $R_{\mathrm{sh}}$, a constant detector parameter, and $I_{\mathrm{b}}$, which is set to a known value at the beginning of every observing schedule. This responsivity estimate is independent of $I$-$V$ information, so it is unaffected by missing or poor-quality $I$-$V$ data. It provides a reference responsivity value that helps identify outlier responsivity values derived from $I$-$V$ data, and provides a calibration factor of last resort when other methods fail. Applying $S_{I_{\mathrm{b}}}$ to calibrate detector TODs has the drawback of introducing a gain bias factor $b_\mathrm{S}$ (see Table~\ref{table:resp}) defined as
\begin{equation}
b_\mathrm{S} = \frac{S_{I_{\mathrm{b}}}}{S}.
\label{resp_bS}
\end{equation}

\subsubsection{$I$-$V$ DC Responsivity}
\label{sec:iv_resp}
For CLASS TES bolometers,  $P_{\upvarphi}$ changes across the transition since the TES temperature $T$ can vary slightly with TES current $I$ and TES resistance $R$.
The TES voltage bias circuit translates the dependence on $R$ into a function of $I$ and $I_{\mathrm{b}}$ (see Equation~\ref{eqn:R_to_Ib}). 

For example, the $I$-$V$ data plotted in Figure~\ref{fig:iv} shows $P_{\mathrm{J}}$ decreasing with increasing $I$. 
The detector optical loading is approximately constant during the $I$-$V$ data acquisition, therefore $\Delta P_{\upvarphi} \approx \Delta P_{\mathrm{J}}$. This means that the plotted $I$-$V$ data also shows a decreasing $P_{\upvarphi}$ with increasing $I$ when biased on transition.
Since  $P_{\upvarphi}$  is not constant along the TES transition, the responsivity value calculated assuming the ideal case described by Equation~\ref{resp_3} is only approximate. 

Assuming $P_{\gamma}$ is constant during the $\sim$\SI{60}{\second} $I_{\mathrm{b}}$ sweep of an $I$-$V$ acquisition, then
\begin{equation}
\Delta P_{\upvarphi}(I,I_{\mathrm{b}}) = \Delta P_{\mathrm{J}}(I,I_{\mathrm{b}}),
\label{resp_dp_dpp}
\end{equation}
\begin{equation}
\Delta P_{\upvarphi}(I,I_{\mathrm{b}}) =  \left.\frac{\partial P_{\upvarphi}}{\partial I}\right|_{I_{\mathrm{b}}} \Delta I +  \left.\frac{\partial P_{\upvarphi}}{\partial I_{\mathrm{b}}}\right|_{I} \Delta I_{\mathrm{b}},
\label{resp_dpp}
\end{equation}
and
\begin{equation}
\Delta P_{\mathrm{J}}(I,I_{\mathrm{b}}) =  \left.\frac{\partial P_{\mathrm{J}}}{\partial I}\right|_{I_{\mathrm{b}}} \Delta I +  \left.\frac{\partial P_{\mathrm{J}}}{\partial I_{\mathrm{b}}}\right|_{I} \Delta I_{\mathrm{b}}.
\label{resp_dp}
\end{equation}
Substituting equations \ref{resp_dp} and \ref{resp_dpp} into \ref{resp_dp_dpp}, and dividing by $\Delta I$ yields
\begin{equation}
     \left.\frac{\partial P_{\upvarphi}}{\partial I} \right|_{I_{\mathrm{b}}}=  \left.\frac{\partial P_{\mathrm{J}}}{\partial I}\right|_{I_{\mathrm{b}}}  +  \left.\frac{\partial P_{\mathrm{J}}}{\partial I_{\mathrm{b}}}\right|_{I} \frac{\Delta I_{\mathrm{b}}}{\Delta I} - \left.\frac{\partial P_{\upvarphi}}{\partial I_{\mathrm{b}}}\right|_{I} \frac{\Delta I_{\mathrm{b}}}{\Delta I}.
\label{resp_dpp_di}
\end{equation}
We can substitute Equation~\ref{resp_dpp_di} for the first term on the right hand side of Equation~\ref{resp_2} to obtain
\begin{equation}
S^{-1}=\left.\frac{\partial P_{\mathrm{J}}}{\partial I_{\mathrm{b}}}\right|_{I}\frac{\Delta I_{\mathrm{b}}}{\Delta I}- \left.\frac{\partial P_{\upvarphi}}{\partial I_{\mathrm{b}}}\right|_{I} \frac{\Delta I_{\mathrm{b}}}{\Delta I}=R_{\mathrm{sh}}I\frac{\Delta I_{\mathrm{b}}}{\Delta I}- \left.\frac{\partial P_{\upvarphi}}{\partial I_{\mathrm{b}}}\right|_{I} \frac{\Delta I_{\mathrm{b}}}{\Delta I}.
\label{resp_5}
\end{equation}

By taking the partial derivative of Equation~\ref{eqn_Pkappa} with respect to $I_{\mathrm{b}}$, we arrive at
\begin{equation}
    \left.\frac{\partial P_{\upvarphi}}{\partial I_{\mathrm{b}}}\right|_{I}
    = n \kappa T^{n-1}\left.\frac{\partial T}{\partial I_{\mathrm{b}}}\right|_{I}
    = G \left.\frac{\partial R}{\partial I_{\mathrm{b}}}\right|_{I} \left.\frac{\partial T}{\partial R}\right|_{I}
    = G \frac{R_{\mathrm{sh}}}{I}\left.\frac{\partial T}{\partial R}\right|_{I},
    \label{eqn_dPphi_dIb}
\end{equation}
where we have used the relationship between $I_{\mathrm{b}}$ and $R$ in Equation~\ref{eqn:R_to_Ib} to change the dependence $T(I,I_\mathrm{b})$ back to the standard $T(I,R)$\footnote{Note the simple chain rule expansion of the partial derivative is applicable as we are changing only one variable}. Combining Equation~\ref{resp_5} and Equation~\ref{eqn_dPphi_dIb} yields
\begin{equation}
    S^{-1}= R_{\mathrm{sh}}I\frac{\Delta I_{\mathrm{b}}}{\Delta I} 
    \left(1-\frac{G}{I^2} \left.\frac{\partial T}{\partial R}\right|_{I}  \right),
    \label{eqn_resp_6}
\end{equation}
or, in terms of TES loop gain $\mathscr{L}_I$,

\begin{equation}
    S^{-1} = R_{\mathrm{sh}}I\frac{\Delta I_{\mathrm{b}}}{\Delta I} 
    \left(1-\mathscr{L}_I^{-1}\right).
    \label{eqn_resp_7}
\end{equation}

Equation~\ref{eqn_resp_7} places any dependence on $\beta$ into the measured $I$-$V$ slope factor.  By comparing Equation~\ref{eqn:resp_ih2} to Equation~\ref{eqn_resp_7}, we find that the slope\footnote{At the operating TES bias current, the slope of the $I$-$V$ curve is equivalent to the amplitude ratio of an applied bias current step over the TES current bias step response. Therefore, additional electrical bias step measurements can reduce the statistical uncertainty on the $I$-$V$ responsivity estimate and/or provide an alternative method to measure responsivity on shorter time scales, as long as the TES absolute current $I$ is known.} of the $I$-$V$ data is related to the $\mathscr{L}_I$ and $\beta$ parameters through
\begin{equation}
    \frac{\Delta I_{\mathrm{b}}}{\Delta I} = 1-\frac{R}{R_{\mathrm{sh}}} \left(\frac{1+\beta+\mathscr{L}_I}{\mathscr{L}_I-1} \right).
    \label{eqn_iv_slope}
\end{equation}

In the limit of $\mathscr{L}_I \gg 1$ (i.e., large $\alpha$):

\begin{equation}
S\approx \frac{1}{R_{\mathrm{sh}}I}\frac{\Delta I}{\Delta I_{\mathrm{b}}}.
\label{resp_1}
\end{equation}

 Four-wire measurements of CLASS TES MoAu bilayer show $\alpha\approx200$. 
 During dark tests with $T_\mathrm{b} \ll T_{\mathrm{c}}$, $P_{\mathrm{J}} \approx P_{\upvarphi} \approx \kappa T_{\mathrm{c}}^n $, hence Equation~\ref{eqn_loop_gain} is approximated by $\mathscr{L}_I \approx (\kappa T_{\mathrm{c}}^n \alpha) /(n \kappa T_{\mathrm{c}}^{n-1} T_{\mathrm{c}})= \alpha/n \approx 200/n = 50$.
The optical loading on the detectors, under typical observing  condition from the CLASS site, reduces the Joule power by up to a half ($P_{\mathrm{J}}\sim P_{\upvarphi}/2$), which implies $\mathscr{L}_I \approx 25$.
 By employing Equation~\ref{resp_1} to calculate the CLASS responsivity, we are introducing a bias of $\sim$4\%.
 This is smaller than the $\sim$20\% bias introduced by using $S^*$ or $S_{I_{\mathrm{b}}}$ responsivity approximations (See $b_{\mathrm{S}}$ in Table~\ref{tab:class_summary}).
 
For the $I$-$V$ responsivity model derived in this subsection, the gain bias defined by Equation~\ref{resp_bS} is
\begin{equation}
    b_\mathrm{S} = \frac{S_{I_{\mathrm{b}}}}{S} = -\frac{I}{I_{\mathrm{b}}} \frac{\Delta I_{\mathrm{b}}}{\Delta I} \left(1-\mathscr{L}_I^{-1}\right)
    \approx -\frac{I}{I_{\mathrm{b}}} \frac{\Delta I_{\mathrm{b}}}{\Delta I}.
\label{resp_bS_IV}
\end{equation}

\subsection{Detector time constant}
\label{sec:iv_analysis_tau}
\begin{figure}
\tikzstyle{block} = [draw, fill=blue!10, rectangle,
    minimum height=3em, minimum width=6em]
\tikzstyle{sum} = [draw, fill=blue!10, circle, node distance=1cm]
\tikzstyle{input} = [coordinate]
\tikzstyle{output} = [coordinate, node distance=3.5cm]
\tikzstyle{pinstyle} = [pin edge={to-,thin,black}]

\begin{tikzpicture}[auto, node distance=2cm,>=latex']
    \node [input, name=input] {};
    \node [sum, right of=input] (sum) {};
    \node [block, right of=sum, node distance = 3.7cm] (controller) {$A \equiv \left.\frac{\partial I}{\partial P_{\upvarphi}}\right|_{I_{\mathrm{b}}} = \Big(\left.\frac{\partial P_{\gamma}}{\partial I}\right|_{I_{\mathrm{b}}}+\left.\frac{\partial P_{\mathrm{J}}}{\partial I}\right|_{I_{\mathrm{b}}}\Big)^{-1}$};
    \node [output, right of=controller] (output) {};
    \node [block, below of=controller] (measurements) {$B \equiv\left.\frac{\partial P_{\mathrm{J}}}{\partial I}\right|_{I_{\mathrm{b}}}= R_{\mathrm{sh}}(I_{\mathrm{b}}-2 I)$};
    \draw [draw,->] (input) -- node {$\Delta P_{\gamma}$} (sum);
    \draw [->] (sum) -- node 
    {${\Delta P_{\upvarphi}}$} (controller);
    \draw [->] (controller) -- node [name=y] {$\Delta I$}(output);
    \draw [->] (y) |- (measurements);
    \draw [->] (measurements) -| node[pos=0.99] {$+$}
        node [near end] {$\Delta P_{\mathrm{J}}$} (sum);
\end{tikzpicture}
\caption{Block diagram of a TES model with negative electrothermal feedback. $A < 0$ because as $I$ increases ($R$ decreases), $T$ is suppressed and $P_{\upvarphi}$ decreases. $B > 0$ since $I_{\mathrm{b}} \gg I$ due to the voltage bias circuit defined by the shunt resistor $R_{\mathrm{sh}} \ll R$. Therefore $AB < 0$, and the feedback is negative.
The gain of the feedback loop is $A/(1-AB) = \left.\partial I/\partial P_{\gamma}\right|_{I_{\mathrm{b}}}$, or $1/S = 1/A-B$, which is equivalent to Equation~\ref{resp_2}.
As $A$ approaches negative infinity in the ideal TES case of a perfectly sharp transition, DC responsivity equals $-1/B$, as expected from Equation~\ref{resp_3}.
}
\label{fig:tes_bol}
\end{figure}
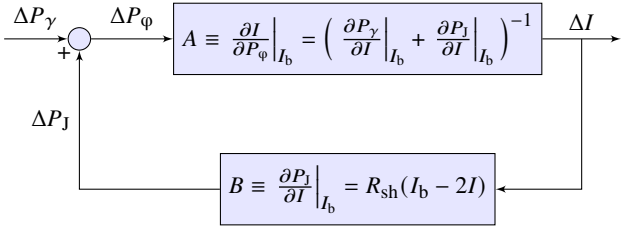
Changes in the TES bolometer optical time constant ($\tau_{\gamma}$) during sky observations  can be tracked through $I$-$V$ measurements.
The bolometer thermal time constant ($\tau$) is the ratio of its heat capacity ($C$) to the thermal conductance ($G$) between the bolometer island and surrounding silicon frame.
The TES electrical time constant ($\tau_{\mathrm{el}}$) is always smaller than the ratio of the TES loop inductance ($L$) and the TES resistance ($R$) since $\beta > 0$.
A typical CLASS TES operates with $\tau_{\mathrm{el}} < L/R = \SI{500}{\nano\henry}/\SI{2.5}{\milli\ohm}=\SI{0.2}{\milli\second}$, much faster than the quickest thermal response (CLASS G-band array)  $\tau = \SI{5}{\pico\joule\per\kelvin}/\SI{808}{\pico\watt\per\kelvin} \approx~\SI{6}{\milli\second} $.
In this case, the electrothermal feedback loop created by voltage-biasing the TES can be considered to have infinite bandwidth, since $\tau_{\mathrm{el}} \ll \tau$. 
A feedback system as shown in Figure~\ref{fig:tes_bol} with instantaneous feedback response (see Appendix~\ref{appendixA}) speeds up $\tau_{\gamma}$ by a factor:

\begin{equation}
\frac{\tau_{\gamma}}{\tau} = \frac{1}{1- A B} = 1-\frac{S}{S^*} = 1+\left(\frac{I_{\mathrm{b}}}{I}-2\right)\frac{\Delta I}{\Delta I_{\mathrm{b}}}\left(1-\mathscr{L}_I^{-1}\right)^{-1},
\label{eqn:tau_factor}
\end{equation}
where
\begin{equation}
A^{-1} = \left.\frac{\partial P_{\gamma}}{\partial I}\right|_{I_{\mathrm{b}}}+\left.\frac{\partial P_{\mathrm{J}}}{\partial I}\right|_{I_{\mathrm{b}}},\mathrm{and}
\end{equation}
\begin{equation}
 B =\left.\frac{\partial P_{\mathrm{J}}}{\partial I}\right|_{I_{\mathrm{b}}}. 
\end{equation}
In the limit of high loop gain ($\mathscr{L}_I \gg 1$),  Equation~\ref{eqn:tau_factor} allows us to compute the speed-up factor between the optical time constant $\tau_{\gamma}$ and the TES thermal time constant $\tau$ from $I$-$V$ data alone. Alternatively, we can constrain $\tau$ by combining $I$-$V$ data with an independent measurements of $\tau_{\gamma}$\footnote{ With the TES biased at the $I$-$V$ inflection point where $\Delta I/\Delta I_{\mathrm{b}} = 0$ and $\mathscr{L}_I = 1$, Equation~\ref{eqn:tau_factor} turns into
$\tau_{\gamma}/\tau = 1+\left(R_{\mathrm{sh}}/R-1\right)\left(1/(2+\beta)\right) \approx 1/2.$
At this bias setting high on the transition $\beta\approx0$~\citep{ullom_review_2015}, and we can constrain $\tau$ by measuring $\tau_{\gamma}$ with bias steps or other methods.}.

Using Equation~\ref{eqn:R_to_Ib} and Equation~\ref{eqn_iv_slope} to express Equation~\ref{eqn:tau_factor} $\tau_{\gamma}/\tau$ in terms of $R$, $R_{\mathrm{sh}}$, $\mathscr{L}_I$ and $\beta$ yields
\begin{multline}
\frac{\tau_{\gamma}}{\tau} =1 + \left[\left(\frac{R}{R_{\mathrm{sh}}}-1\right)
\left(-\frac{R}{R_{\mathrm{sh}}}\frac{1}{\mathscr{L}_I-1}\right)^{-1}\right.
\\\left.\left( 1+\beta+\frac{R_{\mathrm{sh}}}{R}+\left(1-\frac{R_{\mathrm{sh}}}{R}\right)\mathscr{L}_I \right)^{-1}
\left(1-\mathscr{L}_I^{-1}\right)^{-1}\right],
\label{eqn:tau_eff_1}
\end{multline}
\begin{equation}
\begin{split}
\frac{\tau_{\gamma}}{\tau} = \frac{1+\beta+R_{\mathrm{sh}}/R}
{1+\beta+R_{\mathrm{sh}}/R+\left(1-R_{\mathrm{sh}}/R\right)\mathscr{L}_I}
\equiv \frac{\tau_{\mathrm{eff}}}{\tau}.
\end{split}
\label{eqn:tau_eff_2}
\end{equation}
This definition for $\tau_{\mathrm{eff}}$ matches the one presented in \citep{irwin_hilton}, and its interpretation as the detector optical time constant in the limit of low circuit inductance (i.e., instantaneous TES electrical circuit response).

While observing the sky, the typical CLASS TES optical time constant is around five times faster than the intrinsic detector thermal time constant ($\tau_{\gamma}/\tau\sim 0.2$). 
In addition to correcting the detector gain by the TES DC responsivity, we divide out a single-pole transfer function that corrects changes in gain and phase across the signal band ($\sim$\SI{10}{\hertz}) due to the detector optical time constant.
We obtain very accurate time constant measurements from analysing the VPM synchronous signal (VSS)~\citep{katie_thesis}, which are cross-checked with the $I$-$V$ derived time constant estimate described in this section.
The detector time response is not a part of the DC calibration considered in Sections~\ref{sec:class_calib} and \ref{sec:calib_test}.

\subsection{TES stability}
\label{sec:iv_analysis_stability}
The TES electrothermal feedback is stable as long as ${\tau_{\mathrm{el}} (\mathscr{L}_I-1) < \tau}$~\citep{irwin_hilton} or equivalently
\begin{equation}
\frac{LG}{RC} < \frac{R_{\mathrm{sh}}/R+1+\beta}{\mathscr{L}_I-1} < \frac{2+\beta}{\mathscr{L}_I-1}~\mathrm{for}~ R_{\mathrm{sh}} < R. 
\label{eqn:tes_stability}
\end{equation}
As the detector is biased lower on the transition, this condition is harder to satisfy since $R$ decreases. 
To avoid TES instability, CLASS aims to operate all bolometers above 30\% of $R_{\mathrm{n}}$. 
On the other hand, operating the TES lower on the transition has the benefit of suppressing detector Johnson noise through electrothermal feedback and reducing the noise contribution of the SQUID readout. 
The uniformity of the Q-band and G-band TES parameters~\citep{dchuss_class_uniformity,sumit_hf} allows all detectors in these arrays to be biased between 30\% and 60\% $R_{\mathrm{n}}$.
This has translated to optimized sensitivity and robust detector stability across a wide range of operating conditions.
The $T_{\mathrm{c}}$ variations across W-band detector wafers and the susceptibility of W-band TES to instability at higher \%$R_{\mathrm{n}}$ compared to the other arrays limit the number of  W-band detectors biased on transition at the same time to $\sim$80\% of the array~\citep{sumit_4year_det}.

We can use Equation~\ref{eqn:R_to_Ib} to replace $R/R_{\mathrm{sh}}$ in Equation~\ref{eqn_iv_slope} to obtain
\begin{equation}
    \frac{\mathscr{L}_I-1}{\beta+2} = \frac{I_{\mathrm{b}}-I}{2I-I_{\mathrm{b}}-I \Delta I_{\mathrm{b}}/\Delta I}.
\end{equation}

The CLASS Q-band TES film design later replicated in the  W-band and G-band arrays was guided by estimates of the ratio of $\mathscr{L}_I$ (or $\alpha$) to $\beta$ extracted from $I$-$V$ data~\citep{spie_jappel}. 
The chosen TES architecture typically operates with $(\beta+2)/(\mathscr{L}_I-1) > 0.2$ (see Table~\ref{tab:class_summary}), and a ratio $LG/RC < (\SI{500}{\nano\henry}/\SI{2.5}{\milli\ohm}) ~(\SI{808}{\pico\watt\per\kelvin}/\SI{5}{\pico\joule\per\kelvin}) = 0.032$, therefore comfortably satisfying the stability condition  of Equation~\ref{eqn:tes_stability}. 

\subsection{CLASS TES model assumptions and approximations}
Here we discuss the assumptions and approximations of the described TES bolometer model in the context of the CLASS detectors.
We are interested in the TES operation at equilibrium, hence we focus on the detector
DC responsivity and model the temporal response with a single-pole transfer function given by the optical time constant. We find this approach adequately captures the behaviour of the CLASS TES bolometers. 

The CLASS detectors operate in the low-inductance limit and hence the electrical time constant is much faster than the thermal time constant. As noted in section~\ref{sec:iv_analysis_tau} the electrical time constant is $\sim$\SI{0.2}{\milli\second} and the fastest CLASS thermal time constant is $\sim$\SI{6}{\milli\second}.
Equation~\ref{eqn:resp_ih2} indicates that the electrical-inductance $L$ of the TES circuit does not effect the TES DC responsivity, hence moving away from the low-inductance limit predominantly affects the temporal response of the TES.

The responsivity Equation~\ref{eqn_resp_7} relies on making the linear approximations described by Equations~\ref{resp_dpp} and \ref{resp_dp}, hence it is only valid on the small signal limit, where $\Delta P_{\gamma} \ll P_{\mathrm{sat}}$. For CLASS, this condition fails when observing bright sources like the Sun, or the Moon with the \SIlist{90;150;220}{\giga\hertz} arrays, and when cloudy and/or high PWV atmospheric conditions appear at the site, in particular for the \SI{220}{\giga\hertz} receiver.

For CLASS, we calculate the detector responsivity from Equation~\ref{resp_1}, which is a good approximation in the high $\mathscr{L}_I$ limit. This is typically the case when the detectors operated in the middle-to-low resistance range of the TES transition. Detectors operating high on the transition ($>70\% R_{\mathrm{n}}$) have low enough $\mathscr{L}_I$ to introduce significant bias to the DC responsivity estimate.

When acquiring $I$-$V$ data in the field, the optical signal is not perfectly constant. Typically, the atmospheric conditions do not change much during the one minute $I$-$V$, but during poor weather conditions the change can be significant.  Additionally, we acquire $I$-$V$s with the telescope mount scanning and the VPM running. This means the $I$-$V$ data is susceptible to the VPM synchronous signal (VSS). In the next section, we show that the $I$-$V$ bin responsivity estimate helps average down the systematic error caused by superimposing the small VSS signal on the $I$-$V$ data.
This systematic error averages down because the phase of the VSS is random between different $I$-$V$ data sets, and the VSS \SI{10}{\hertz} frequency is fast compared to the $I_{\mathrm{b}}$ rate of change during the $I$-$V$.  

\section{$I$-$V$~bin Calibration method}
\label{sec:iv_analysis_ivbin}
The set of all $I$-$V$ measurements for each detector can be grouped into bins based on $P_{\mathrm{sat}}$, which is a proxy for optical loading ($P_{\gamma} = P_{\upvarphi}-P_{\mathrm{sat}}$).
Analysis of $I$-$V$s grouped in the same bin yield similar detector parameters as long as the bin range is small. 
The CLASS Q-band $I$-$V$ data sets are grouped in bins of 0.1~pW, while for W-band and G-band, the bin widths are 0.2~pW. These bin widths are only a few percent of the total detector $P_{\mathrm{sat}}$.

All raw $I$-$V$ data are calibrated to $I$ versus $I_{\mathrm{b}}$ units and saved in a database indexed by detector number and $P_{\mathrm{sat}}$.
The mean and error estimate of a detector parameter (e.g., responsivity, time constant, etc) is extracted from the database by:  
\begin{enumerate}
\item finding the group of $I$-$V$s with $P_{\mathrm{sat}}$ values in the desired range;
\item for all $I$-$V$ in the $P_{\mathrm{sat}}$ group, computing the parameter value across a small $I_{\mathrm{b}}$ range centered at the target $I_{\mathrm{b}}$;
\item eliminating outliers from the distribution of parameter results; and
\item computing the mean and standard deviation of the distribution.
\end{enumerate}

As observing time on the sky accumulates, more $I$-$V$s are acquired, which increases the number of measurements in each $P_{\mathrm{sat}}$ bin, leading to three advantages. 
First, it improves the precision of the parameter result per bin.
Second, it allows for the recovery of detector parameters in the few instances the $I$-$V$ acquisition failed, but the detector behaved properly during observations, by assuming the bias parameters of the otherwise-well-behaved detector are equal to bin averages.
Third, it populates bins over a wide range of $P_{\mathrm{sat}}$, revealing detector behavior beyond the standard small-signal TES bolometer model.
Applying the $I$-$V$~bin method to Q-band data acquired between June 2016 and August 2020 yields a responsivity median standard error of 0.3\%.

\begin{figure*}
\begin{center}
    \includegraphics[width=0.8\linewidth]{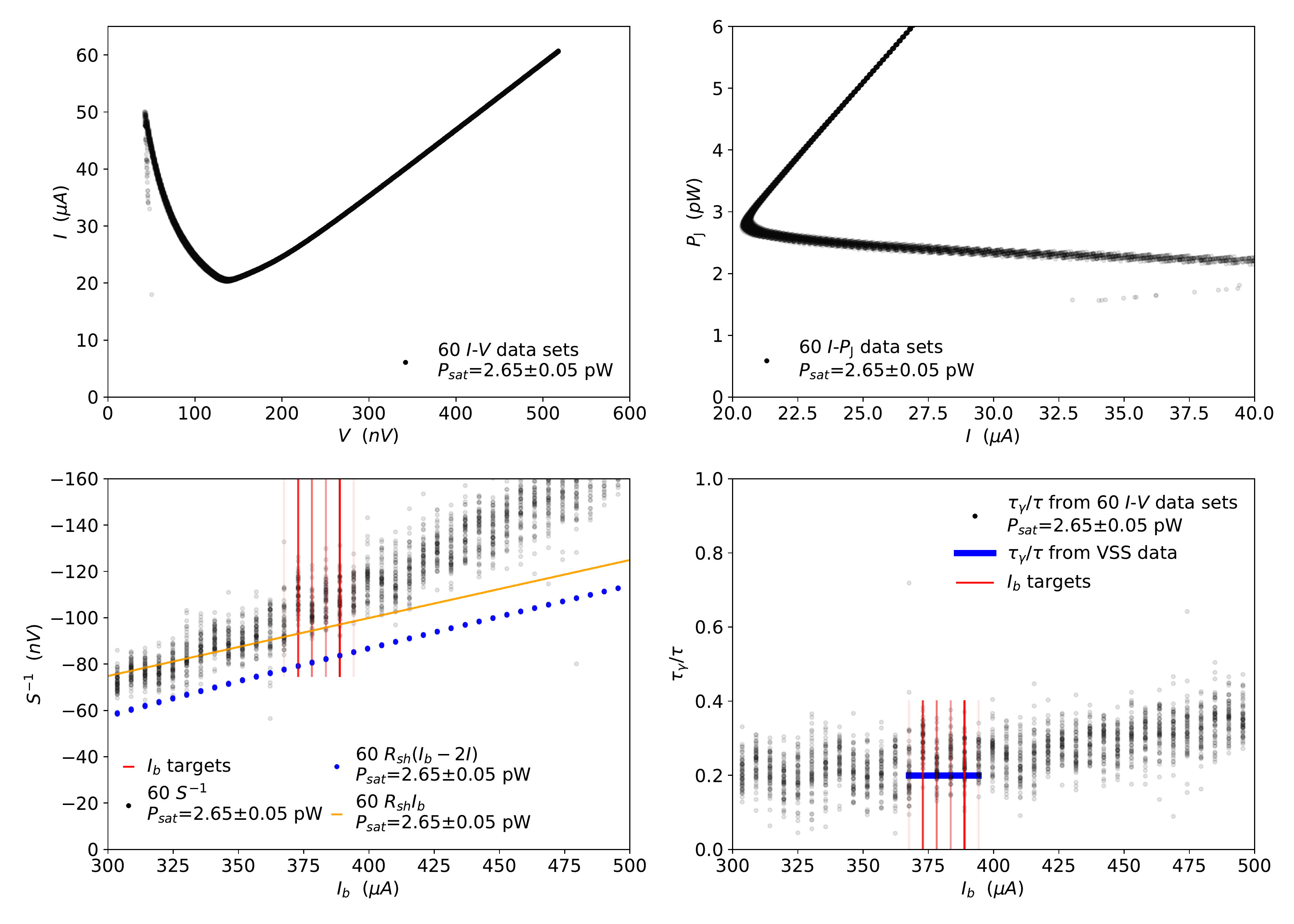}
\end{center}
\caption{Example of the $I$-$V$~bin analysis method applied to data from a CLASS Q-band TES.
\textbf{Top left}: The black dots plot 60 $I$-$V$ data sets from the same detector, acquired under almost identical optical loading conditions with the detector saturation power $P_{\mathrm{sat}}$ constrained to \SI{2.65\pm0.05}{\pico\watt}. The 60 $I$-$V$ data sets were acquired between June 2016 and March 2018. Notice that the overlap between the $I$-$V$s is almost perfect, except at low TES voltage ($V$) where the detector becomes unstable and moves off its transition to become fully superconducting. The reproducibility of the $I$-$V$ curves based on their measured $P_{\mathrm{sat}}$ allows us to group them together in bins. We improve the precision of the derived detector parameters by averaging the results across the bin. Having a distribution of equivalent $I$-$V$ curves simplifies the process of identifying outliers, replacing parameters derived from poor-quality $I$-$V$ sweeps, and estimating the uncertainty of each derived detector parameter.
\textbf{Top right}: The black points show the 60 $I$-$V$ data sets from the top-left plot, but plotted in terms of bias power $P_{\mathrm{J}}$ on the y-axis and TES current ($I$) on the x-axis.
\textbf{Bottom left}: The black points plot the measured $I$-$V$ responsivity (see Section~\ref{sec:iv_resp} and Equation~\ref{resp_1}) for each $I$-$V$ in the $P_{\mathrm{sat}}=\SI{2.65\pm0.05}{\pico\watt}$  bin shown in the top-left figure. The responsivity is plotted on the y-axis as a function of applied bias current ($I_{\mathrm{b}}$) on  the x-axis. The red vertical lines show the target $I_{\mathrm{b}}$ chosen for each of the 60 observation scans following each $I$-$V$. As expected, the target $I_{\mathrm{b}}$ chosen for $I$-$V$ sweeps in the same $P_{\mathrm{sat}}$ bin are similar, validating the algorithm used to set the optimal bias target for each column of detectors.
The blue points plot the ideal TES $I$-$V$ responsivity, while the orange line plots the $I_{\mathrm{b}}$ responsivity.
\textbf{Bottom right}: On the y-axis, the black points plot the $\tau_{\gamma}/\tau$ ratio derived from the $I$-$V$s in the $P_{\mathrm{sat}}=$\SI{2.65\pm0.05}{\pico\watt} bin. The ratio is plotted as a function of $I_{\mathrm{b}}$ on the x-axis. The optical time constant $\tau_{\gamma}$ slows down as the applied $I_{\mathrm{b}}$ increases and the TES equilibrium temperature moves higher on the superconducting transition. Higher on the transition, the electrothermal feedback is weaker, causing the slowdown of the TES response. 
The blue line plots the measured $\tau_{\gamma}/\tau$ ratio based on the time constant extracted from the VPM synchronous signal (VSS)~\citep{katie_thesis}.
}
\label{fig:bin_iv}
\end{figure*}

Figure~\ref{fig:bin_iv} shows an example of this method, constraining a detector's responsivity and time constant ratio $\tau_{\gamma}/\tau$ in one $P_{\mathrm{sat}}$ bin across a range of $I_{\mathrm{b}}$ targets.

\section{CLASS detector calibration}\label{sec:class_calib}

Raw CLASS detector data are calibrated to thermodynamic CMB temperature units in five steps described in the following subsections: (1) The raw data are converted to TES current units, which allows the TES signal to be interpreted in the context of the TES model presented in Section~\ref{sec:model}. (2) The responsivity factor derived from the TES model transforms the measured TES current to optical power deposited at the bolometer. This step places on the same footing all the data of a single detector, accounting for any time-dependent variation related to the TES itself, such as changes in applied bias current or optical loading. 
(3) Apply gain calibration factors between detectors, thereby normalizing the data from the entire array of detectors to the same scale. These gain factors are equivalent to the relative optical efficiency between detectors. Any changes to detector optical efficiency over time (not captured by the responsivity calibration) can be incorporated at this step.
(4) Apply a per-detector atmospheric opacity correction based on detector frequency band, elevation pointing, and precipitable water vapor (PWV) during the observation. The atmospheric opacity model is described in Section~\ref{sec:atm}. 
(5) The array-normalized power is converted to thermodynamic temperature by calibrating off a bright Rayleigh\textendash{Jeans} source such as the Moon \citep{jappel_Q, zxu_Q} or a planet \citep{sumit_4year_det}, and applying a conversion factor from Rayleigh\textendash{Jeans} temperature to CMB thermodynamic temperature based on the detector bandpass. Alternatively, this last calibration factor can be obtained from cross-correlating CLASS maps constructed in power units with Wilkinson Microwave Anisotropy Probe (WMAP; \citealt{bennett_2013}) and \textit{Planck}~\citep{planck_2018_cosmo_param} maps calibrated to CMB thermodynamic temperature.

In summary, a small $\Delta I_j$ signal of detector $j$ is calibrated to $\Delta T_{\mathrm{cmb}}$ units through
\begin{equation}
    \Delta T_{\mathrm{cmb}} = \underbrace{\Delta I_j}_{\ref{sec:icalib}} \underbrace{S^{-1}_{jk}}_{\ref{sec:ivcalib}}
    \underbrace{\epsilon_j^{-1}}_{\ref{sec:relgain}}
        \underbrace{(1-\mu_{jk}^{\mathrm{atm}})^{-1}}_{\ref{sec:atm}}
    \underbrace{\frac{\mathrm{d}T_{\mathrm{cmb}}}{\mathrm{d}T_{\mathrm{RJ}}}\frac{\mathrm{d}T_{\mathrm{RJ}}}{\mathrm{d}P_\gamma}}_{\ref{sec:tcmbcalib}} ,
\label{eqn:class_tes_calib}
\end{equation}
where $k$ identifies the $I$-$V$ used to compute the detector responsivity $S_{jk}$. Individual detector data are calibrated to array standard power through a relative detector gain factor $\epsilon_j$. The atmospheric opacity gain correction $\mu_{jk}^{\mathrm{atm}}$ is modeled for each detector $j$ at $I$-$V$ time $k$. The power to $T_{\mathrm{cmb}}$ conversion factor $\left.\mathrm{d}T_{\mathrm{cmb}}/\mathrm{d}T_{\mathrm{RJ}}\right. \left.\mathrm{d}T_{\mathrm{RJ}}/\mathrm{d}P_\gamma\right.$ is a single value for each array. 
Each term in Equation~\ref{eqn:class_tes_calib} is labeled from below with the section number discussing that calibration step.

\subsection{Calibration to TES current}\label{sec:icalib}

\begin{deluxetable}{ccccc}[ht]
\tablenum{1}
\tablehead{
\colhead{} & \colhead{\textbf{40 GHz}} & \colhead{\textbf{90 GHz}} & \colhead{\textbf{150 GHz}} & \colhead{\textbf{220 GHz}}
}
\startdata
    $R_{\mathrm{FB}}$ (\SI{}{\ohm}) & 5100 & 2100 & 2100 & 2100 \\
    $V_{\mathrm{FB}}$ (\SI{}{\volt}) & 1 & 1 & 1 & 1 \\
    $\mathrm{FB}_{\mathrm{bits}}$ & 14 & 14 & 14 & 14\\
    $M_\mathrm{r}$ & 24.6 & 24.6 & 24.6 & 24.6 \\
    $B_\mathrm{g}$ & 2048 & 2048 & 2048 & 2048 \\
    $R_{\mathrm{b}}$ (\SI{}{\ohm}) & 573 & 573 & 573 & 573 \\
    $V_{\mathrm{b}}$ (\SI{}{\volt}) & 5 & 5 & 5 & 5  \\
    $\mathrm{b}_{\mathrm{bits}}$ & 15 & 15 & 15 & 15 \\
    $R_{\mathrm{sh}}$ (\SI{}{\micro\ohm}) & 250 & 250 & 200 & 200\\
\enddata
\caption{CLASS MCE readout and biasing parameters used to calibrate raw data in digital counts to physical units of electrical current.  }
\label{tab:class_mce}
\end{deluxetable}

The receiver data are recorded by a Multi-Channel Electronics (MCE) box developed at the University of British Columbia \citep{ubc_mce}. 
The MCE records changes in the digital-to-analog (DAC) feedback ($\Delta_{\mathrm{FB}}$) applied to the readout loop that nulls the signal of each channel.
The TOD of this feedback signal represents the TES response to the sky signal.
We first convert the TOD from raw DAC counts to TES current ($I$) through
\begin{equation}    
\Delta I = \Delta_{\mathrm{FB}}\frac{V_{\mathrm{FB}}}{2^{\mathrm{FB}_{\mathrm{bits}}}}\frac{1}{R_{\mathrm{FB}}}\frac{1}{M_\mathrm{r}}\frac{1}{B_\mathrm{g}},
\label{eqn:dac_I}
\end{equation}
where $V_{\mathrm{FB}}$ is the maximum voltage of the DAC, $\mathrm{FB}_{\mathrm{bits}}$ the resolution of the feedback DAC, $R_{\mathrm{FB}}$ the total resistance of the feedback circuit, $M_\mathrm{r}$ the mutual inductance ratio between the TES and SQUID feedback coupling coils \citep{nist_tdm_mux13b}, and $B_\mathrm{g}$ the DC gain of the 4-pole Butterworth filter applied by the MCE before down-sampling to record the data at $\sim$\SI{201}{\hertz}. 
The Butterworth filter is eventually divided out from the raw data to recover the TES response across the entire sampling bandwidth.
Table~\ref{tab:class_mce} summarizes the CLASS MCE calibration parameter values.

The MCE applies the bias current ($I_{\mathrm{b}}$) used to turn on the TES detectors. $I_{\mathrm{b}}$ is calibrated from digital counts ($\mathrm{DAC_b}$) to amperes through
\begin{equation}    
I_{\mathrm{b}} = \mathrm{DAC_b}\frac{V_{\mathrm{b}}}{2^{\mathrm{b}_{\mathrm{bits}}}}\frac{1}{R_{\mathrm{b}}},
\label{eqn:dac_Ib}
\end{equation}
where $V_{\mathrm{b}}$ is the maximum voltage of the detector bias DAC, $\mathrm{b}_{\mathrm{bits}}$ its resolution, and $R_{\mathrm{b}}$ the total resistance of the detector bias circuit.

\subsection{Calibration to power detected at the TES}
\label{sec:ivcalib}
\begin{deluxetable}{ccccc}[ht]
\tablenum{2}
\label{table:resp}
\tablehead{
\colhead{} & \colhead{\textbf{40 GHz}} & \colhead{\textbf{90 GHz}} & \colhead{\textbf{150 GHz}} & \colhead{\textbf{220 GHz}}
}
\startdata
$S^{-1}$ (\SI{}{\nano\volt}) & $-127 (-122)$  & $-336$  & $-422$  & $-392$  \\
$\sigma_\mathrm{S}^\mathrm{a}$ (\%) & 24 (24)  & 24  & 26  & 45  \\
$\sigma_\mathrm{S}^\mathrm{d}$ (\%) & 13 (13)  & 24  & 21  & 25  \\
$N_{IV}$  & 1594 (2276)  & 1508  & 520  & 520  \\
$b_\mathrm{S}$ & 1.09 (1.09)  & 1.29  & 1.17  & 1.22  \\
$\sigma_{b_\mathrm{S}}$ (\%)  & 6 (6)  & 9  & 6  & 8  \\
$\tau_{\gamma}/\tau$  & 0.21 (0.20)  & 0.29  & 0.22  & 0.22 \\
$(\mathscr{L}_I-1)/(\beta+2)$  & 4.1 &  2.4 & 4.6  & 4.1  \\
$\sigma_{\epsilon_j}$ (\%) & 6  & 45  & 18  & 18  \\
$\eta$ & 0.54 (0.43) & 0.42 & 0.45 & 0.45 \\
$\begin{tabular}{@{}c@{}} $dT_{\mathrm{RJ}}/dP_{\gamma}$ \\ $(\SI[per-mode = symbol]{}{\kelvin\per\pico\watt})$ \end{tabular}$ & 10.8 (13.8)  & 5.5  & 5.1  & 4.4  \\
$dT_{\mathrm{cmb}}/dT_{\mathrm{RJ}}$ & 1.04  & 1.23  & 1.68  & 2.92  \\
\enddata
\caption{CLASS detector calibration parameters for data acquired between May 2018 and August 2020.  The \SI{40}{\giga\hertz} parameters show two values, one for the telescope in its nominal configuration and in parentheses for data acquired with a thin-grill filter installed on the cryostat window.
$S$ is the average TES responsivity across the array. 
$\sigma_\mathrm{S}^\mathrm{a}$ is the 1-$\sigma$ responsivity standard deviation across the array.
$\sigma_\mathrm{S}^\mathrm{d}$ is the 1-$\sigma$ standard deviation of the per-detector normalized responsivity.
$N_{IV}$ is the number of $I$-$V$ measurements acquired.
${b_\mathrm{S}}$ is the array average bias factor between $I$-$V$~bin responsivity and $I_\mathrm{b}$ responsivity.
$\sigma_{b_\mathrm{S}}$ is 1-$\sigma$ standard deviation of the array responsivity bias factor.
$\tau_{\gamma}/\tau$ is the array average electrothermal speed-up factor.
The array average ratio of TES parameters $(\mathscr{L}_I-1)/(\beta+2)$ is estimated from $I$-$V$ data.
$\sigma_{\epsilon_j}$ is the 1-$\sigma$ standard deviation of the detector relative optical efficiencies.
$\eta$ is the array average optical efficiency.
$dT_{\mathrm{RJ}}/dP_{\gamma}$ is the calibration factor to convert power measured at the bolometers to Rayleigh\textendash{Jeans} temperature ($T_{\mathrm{RJ}}$) on the sky.
$dT_{\mathrm{cmb}}/dT_{\mathrm{RJ}}$ is the conversion factor from sky Rayleigh\textendash{Jeans} temperature to CMB thermodynamic temperature.
}
\label{tab:class_summary}
\end{deluxetable}

At the beginning of every observing schedule, typically once per day~\citep{petroff_class_control}, $I$-$V$ data are acquired and analyzed to find the optimal detector bias voltage that will turn on and place all TES bolometers on their superconducting transition between 30\% and 60\% of their normal resistance ($R_\mathrm{n}$).
From the applied bias current and this single $I$-$V$ curve, we immediately generate responsivity calibration factors using Equation~\ref{resp_1} that are applicable to the subsequent TOD sets.  The statistical error of each responsivity estimate is reduced from 3.0\% to 0.5\% (see section~\ref{sec:moon_calib_test} and Figure~\ref{fig:moon_calib}) by combining $I$-$V$ data acquired between June 2016 and August 2020 under similar optical loading conditions as described in Section~\ref{sec:iv_analysis_ivbin}. 
We divide out a single-pole transfer function to correct for optical time constant gain and phase changes across the signal band. The detector optical time constants are measured from both the VPM synchronous signal and the detector $I$-$V$ curves~(see Section \ref{sec:iv_analysis_tau}).
Table~\ref{table:resp} reports the average responsivity across each array, as well as the array 1-$\sigma$ variance ($\sigma^\mathrm{a}_\mathrm{S}$), and the per-detector normalized (i.e., $\overline{S}_j = 1$) 1-$\sigma$ variance ($\sigma^\mathrm{d}_\mathrm{S}$). 
Following Equation~\ref{resp_bS}, we report the bias factor $b_\mathrm{S}$ and 1-$\sigma$ variance ($\sigma_{b_\mathrm{S}}$), computed from the ratio of $I_{\mathrm{b}}$ responsivity to $I$-$V$ responsivity.

\subsection{Gain calibration between detectors}
\label{sec:relgain}
To place all detector TODs on the same power scale, we divide the TES current signal $\Delta I$ by the $I$-$V$ responsivity (Section~\ref{sec:ivcalib}) and divide by the relative calibration $\epsilon_j$ of detector $j$.
The relative calibration  between detectors is constructed from high signal-to-noise measurements of the Moon and/or planets.  
The $\epsilon_j$ gain factors are equivalent to the relative optical efficiency of each detector, and by construction the mean of all $\epsilon_j$ across an array is equal to 1.
With the TODs for all detectors in the array calibrated to the same power scale, the data can be combined to generate maps in standardized array power units. 
Table \ref{table:resp} reports the standard deviation of relative detector gains ($\sigma_{\epsilon_j}$) across each array. Note that the statistical error on each detector $\epsilon_j$ extracted from point-source observations is <1\%. The $\epsilon_j$ variations observed across the arrays are larger ($\sigma_{\epsilon_j} > 8$\%) than the per-detector measurement uncertainty.

\subsection{Atmospheric opacity correction}
\begin{figure}
    \includegraphics[width=1.0\linewidth]{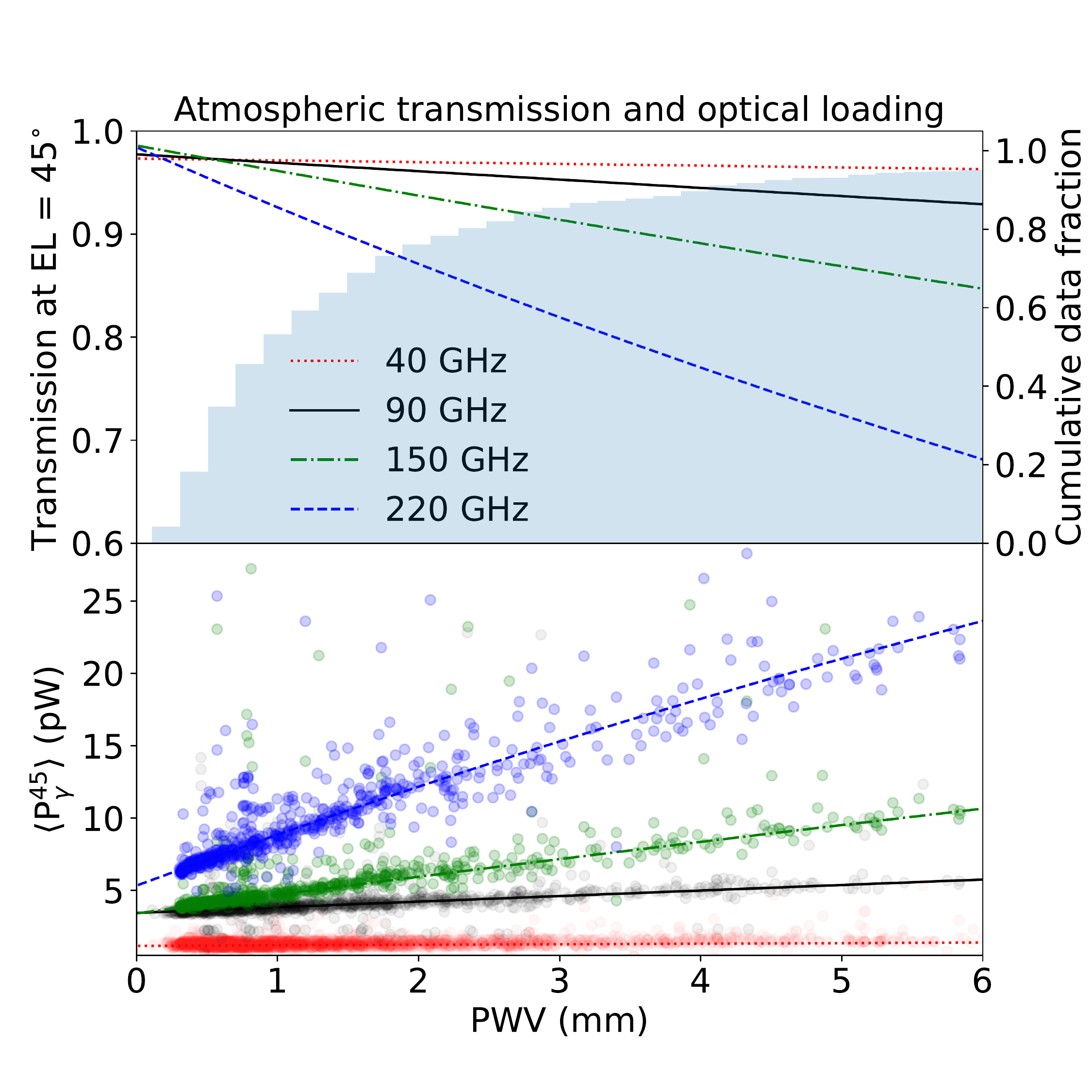}
    \caption{The top plot shows the atmospheric transmission model for each of the CLASS bands (\textcolor{red}{$\bullet$}\SI{40}{\giga\hertz}, \textcolor{black}{$\bullet$}\SI{90}{\giga\hertz}, \textcolor[rgb]{0.13,0.55,0.13}{$\bullet$}\SI{150}{\giga\hertz}, and \textcolor{blue}{$\bullet$}\SI{220}{\giga\hertz}) as a function of PWV, based on \citet{pardo_atm_model}. The shaded blue histogram indicates the cumulative fraction of CLASS CMB scans acquired below a particular PWV value, with about 80\% of data acquired below PWV of \SI{3}{\milli\meter}.
    The lines plotted on the bottom axes indicate the expected optical loading at \SI{45}{\degree} elevation for each CLASS band based on: (1) the transmission model shown on the top plot; (2) an atmospheric temperature of \SI{266}{\kelvin}; (3) receiver optical loading offsets of \SIlist{0.6;2.4;2.7;4.4}{\pico\watt} at \SIlist{40;90;150;220}{\giga\hertz}, respectively; and (4) the brightness temperature to power calibration factors presented in \citep{sumit_4year_det}. Each data point is the averaged detector optical loading in a CLASS band normalized to \SI{45}{\degree} elevation. The per-detector optical loading is derived from $I$-$V$ measurements. The atmospheric transmission model optical loading predictions match well with the $I$-$V$ optical loading measurements across the CLASS bands. }
\label{fig:atm}
\end{figure}

The opacity of the atmosphere along the line-of-sight suppresses the amplitude of the celestial sky signal. 
We apply an atmospheric opacity model based on \citet{pardo_atm_model} to correct the detector gain calibration across the four CLASS frequency bands.
The opacity correction is a function of each detector's elevation pointing, frequency bandpass, and the atmospheric PWV at the beginning of each scan. 
The PWV value input to the opacity model is either: (1)   
measured by nearby radiometers\footnote{Nearby radiometers include the UdeC and UCSC 183 GHz radiometer located at the Atacama Cosmology Telescope site~\citep{parque_atacama}, and the Atacama Pathfinder Experiment radiometer~\citep{apex, pwv_20years}}, or (2) estimated from CLASS $I$-$V$ optical loading  measurements. 
Combining these two methods allows us to build a complete set of PWV values that encompasses all our CMB observations and to reject outliers by checking for consistency between the multiple PWV estimates. 
Figure~\ref{fig:atm} shows the transmission model for the four CLASS bands, the model's predicted optical loading, and the $I$-$V$ measured optical loading as a function of PWV.
The model captures the dependence of the detector optical power on frequency band and PWV, increasing our confidence in the model's atmospheric opacity correction. 
Note that at \SI{40}{\giga\hertz}, the atmospheric transmission is nearly constant up to \SI{6}{\milli\meter} PWV, while the transmission drops by 30\% at \SI{220}{\giga\hertz} across the PWV range.

\label{sec:atm}

\subsection{Calibration to CMB thermodynamic temperature}
\label{sec:tcmbcalib}
With the TODs of all detectors in the array calibrated to the same power standard, the entire data set can be co-added into one map.
The final calibration step consists of converting this co-added map from power units to CMB thermodynamic temperature $T_{\mathrm{cmb}}$. 
This can be achieved by cross-correlating the map with WMAP or \textit{Planck} maps.
Alternatively, we can also estimate this absolute calibration factor from observations of the Moon~\citep{jappel_Q,zxu_Q} and planets \citep{sumit_4year_det, venus_class}. 

The power to $T_{\mathrm{cmb}}$ conversion factor $\frac{\mathrm{d}T_{\mathrm{cmb}}}{\mathrm{d}T_{\mathrm{RJ}}}\frac{\mathrm{d}T_{\mathrm{RJ}}}{\mathrm{d}P_\gamma}$ is a single value for each array where 
\begin{equation}
    \frac{\mathrm{d}T_{\mathrm{RJ}}}{\mathrm{d}P_\gamma} \approx (\eta k_\mathrm{B}\Delta\nu)^{-1},  
\end{equation}
and,
\begin{equation}
    \frac{\mathrm{d}T_{\mathrm{cmb}}}{\mathrm{d}T_{\mathrm{RJ}}}\approx\frac{(e^{x_\mathrm{o}}-1)^2}{x_\mathrm{o}^2 e^{x_\mathrm{o}}}, \quad x_\mathrm{o} = \frac{h \nu_\mathrm{o}}{k_\mathrm{B} T_{\mathrm{cmb}}}. 
\end{equation}
$k_\mathrm{B}$ is Boltzmann's constant, $h$ is Planck's constant, $T_{\mathrm{cmb}}$ is the CMB temperature of \SI{2.725}{\kelvin}~\citep{firas_2009}, $\Delta \nu$ the array average detector optical bandwidth, $\nu_o$ the array average bandpass center frequency, and $\eta$ the array average optical efficiency.

 The power to CMB temperature calibration factors for all four CLASS frequency bands are found in Table~\ref{table:resp}. These values are derived from the bandpass and Rayleigh\textendash{Jeans} source measurements described in \citep{sumit_4year_det}.
 
\subsection{Observing and calibration time scales for CLASS}

CLASS TES bolometer data is sampled every $\sim\SI{5}{\milli\second}$.
The VPM completes a single modulation cycle every $\sim\SI{100}{\milli\second}$. 
The detector optical time constant $\tau_{\gamma}$ is $\sim\SI{2}{\milli\second}$ and is measured to high accuracy by tracking the phase of the VPM synchronous signal~\citep{katie_thesis}.
Corrections of the finite-time response of the detector, including variations in the detector time constant, are not part of the DC calibration described here and will be treated in future work.
The nominal CLASS observing schedule is one day long, and it starts and ends with an $I$-$V$ data acquisition.
The opacity correction method described in section~\ref{sec:atm} is applied per observing schedule. 
We find this approach adequate at the lower CLASS frequency bands due to the relatively small change in atmospheric transmission with a typical change in PWV, but it may require further enhancement, especially  for  the \SI{220}{\giga\hertz} band that is most sensitive to atmospheric water vapor fluctuations. 
We can track atmospheric opacity on shorter time scales by measuring the baseline drifts of the raw data (i.e., the intensity signal) and/or the changes in the VPM synchronous signal amplitude.
Discussion of these methods is left for future publications.

$I$-$V$ data gathered over a period of time where the detector focal plane configuration is unchanged can be grouped to generate an $I$-$V$ bin calibration set. 
Changes to the telescope's cryogenic receivers are rare, typically dividing observations into multi-year periods.
Relative gain calibration periods coincide with changes to the cryogenic receiver or with changes to the \SI{300}{\kelvin} optics, such as the addition of a thin-grill filter to the Q-band telescope window in 2019.
Each subset of data with a relative calibration solution also has a corresponding absolute calibration to CMB thermodynamic temperature. 
A summary of the timescales of CLASS data is found in Table~\ref{tab:class_time_scale}.

\begin{deluxetable}{cc}
\tablenum{3}
\tablehead{
\colhead{} & \colhead{\textbf{Time scale}}}
\startdata
Data sampling rate  & $\sim\SI{5}{\milli\second}$ \\
VPM modulation period  & $\sim\SI{100}{\milli\second}$ \\
Detector optical time constant $\tau_{\gamma}$  & $\sim\SI{2}{\milli\second}$ \\
Detector thermal time constant $\tau$ & $\sim\SI{10}{\milli\second}$ \\
\hline
CLASS CMB scan period  & $\sim\SI{24}{\hour}$ \\
$I$-$V$ calibration period & $\sim\SI{24}{\hour}$ \\
Atmosphere opacity correction & $\sim\SI{24}{\hour}$ \\
\hline
$I$-$V$ bin data sets & $\sim\SI{2}{\year}$ \\
Relative gain calibration between detectors & $\sim\SI{2}{\year}$  \\
Calibration to CMB temperature & $\sim\SI{2}{\year}$ \\
\enddata
\caption{ Time scales of CLASS detector data and calibration steps. 
The detector data sampling rate, detector response time, and VPM modulation are measured in milliseconds. A standard CLASS CMB observing schedule including at least one $I$-$V$ data set is completed in one day. 
The atmospheric opacity gain correction described in section~\ref{sec:atm} is applied per observing schedule. $I$-$V$ data sets are grouped across multiple observing years to generate an optimized $I$-$V$ bin calibration. Absolute and relative detector gain calibration are also computed for multi-year observing periods. }
\label{tab:class_time_scale}
\end{deluxetable}

\section{CLASS calibration tests}\label{sec:calib_test}
\label{sec:moon_calib_test}

This section discusses two tests of the CLASS TES responsivity model. The first test compiles 208 Moon scans acquired by the Q-band telescope between June 2016 and August 2020.
These high signal-to-noise Moon measurements are used as a photometric standard to constrain the uncertainty of the responsivity models.
The second test compares the expected Noise Equivalent Power (NEP) at each CLASS frequency band to the average measured NEP, applying $I$-$V$~bin or $I_\mathrm{b}$ calibration.

\subsection{Q-band Moon photometric calibration test}

To test the accuracy of the detector responsivity calibration, we would like to map a source with uniform brightness temperature across all observation runs. If the calibration method is accurate, the measured variations in the source's brightness will be small.

The Moon is a good photometric calibration standard for the CLASS Q-band telescope because: (1) it is effectively a point source for the 1.5$^\circ$ full-width half max (FWHM) detector beam~\citep{zxu_Q}, (2) it drives a high signal-to-noise response without saturating the detectors, (3) its brightness temperature variations across time have been well characterized~\citep{linsky_moon_center, moon_calib, changE}, and (4) the CLASS scanning strategy covering a large fraction of the sky leads to hundreds of Moon observations throughout the year.

A Moon antenna temperature model~\citep{jappel_Q, zxu_Q} is used to scale the measured Moon signal to a uniform standard. The Moon model accounts for angular diameter variation due to the Moon's orbit and for surface temperature gradients caused by the Sun illuminating one side of the Moon (see Figure~2 in \citealt{zxu_Q}).

The Moon antenna temperatures measured by each detector are computed using four different calibration methods: 
(1) $I$-$V$~bin constructed by grouping $I$-$V$ data based on optical loading (see Section~\ref{sec:iv_analysis_ivbin}),
(2) $I$-$V$ based on a single $I$-$V$ acquired before the scan (see Section~\ref{sec:iv_resp}), 
(3) $I_{\mathrm{b}}$ based on estimating responsivity from the applied detector bias current (see Equation~\ref{resp_Ib}), 
and (4) DAC based on applying no TES detector responsivity calibration.

For each calibration method, Table~\ref{tab:class_moon_test} presents the standard deviation of: (1) 
the per-detector normalized Moon amplitude ($\sigma_\mathrm{M}^\mathrm{d}$), (2)
the normalized ratio of Moon amplitudes ($\sigma_\mathrm{M}^\mathrm{p}$) between pairs of bolometers that share the same feedhorn divided by $\sqrt{2}$ (to convert pair-ratio calibration uncertainty to per-detector uncertainty), (3)
the detector responsivities across the array for the subset of $I$-$V$ data associated to Moon observations ($\sigma^\mathrm{d}_\mathrm{S}$), 
and (4) the relative gain calibration between detectors ($\sigma^d_{\epsilon_j}$).
Additionally, it includes the derived quantity $\sigma_\mathrm{M}^\mathrm{dM}$ obtained by subtracting in quadrature the estimated Moon model uncertainty of 1.5\% from $\sigma_\mathrm{M}^\mathrm{d}$.

\subsubsection{Variance of per-detector Moon amplitudes $\sigma_\mathrm{M}^\mathrm{d}$}
The $I$-$V$~bin calibration method reduces $\sigma_\mathrm{M}^\mathrm{d}$ by nearly a factor of two compared to the standard $I$-$V$ calibration. Therefore, the $I$-$V$~bin calibration is more accurate, since its results are closer to the ideal scenario of the Moon as a perfect photometric standard ($\sigma_\mathrm{M}^\mathrm{d}$ = 0).

Applying the $I_{\mathrm{b}}$ calibration results in a similar $\sigma_\mathrm{M}^\mathrm{d}$ to the $I$-$V$~bin calibration, we conjecture that this calibration method yields accurate results in the Moon photometric test because of the high uniformity of the Q-band observing conditions and detector properties. 
The $I$-$V$~bin calibration is less susceptible to variation in the observing conditions but incorporates additional uncertainty from noise in the $I$-$V$ data. 
Enhancing the quality of the $I$-$V$ data would further improve the overall accuracy of both the $I$-$V$ and $I$-$V$~bin calibrations.

Each Moon measurement included in Figure~\ref{fig:moon_calib} has high signal-to-noise, with statistical uncertainties $<0.2\%$. The variance of the distributions plotted on the top row are dominated by error in the Moon antenna temperature model and the per-detector calibration uncertainty across time. 

\subsubsection{Variance of detector pair-ratio Moon amplitudes $\sigma_\mathrm{M}^\mathrm{p}$}
\begin{figure*}
\begin{center}
    \includegraphics[width=1.00\linewidth]{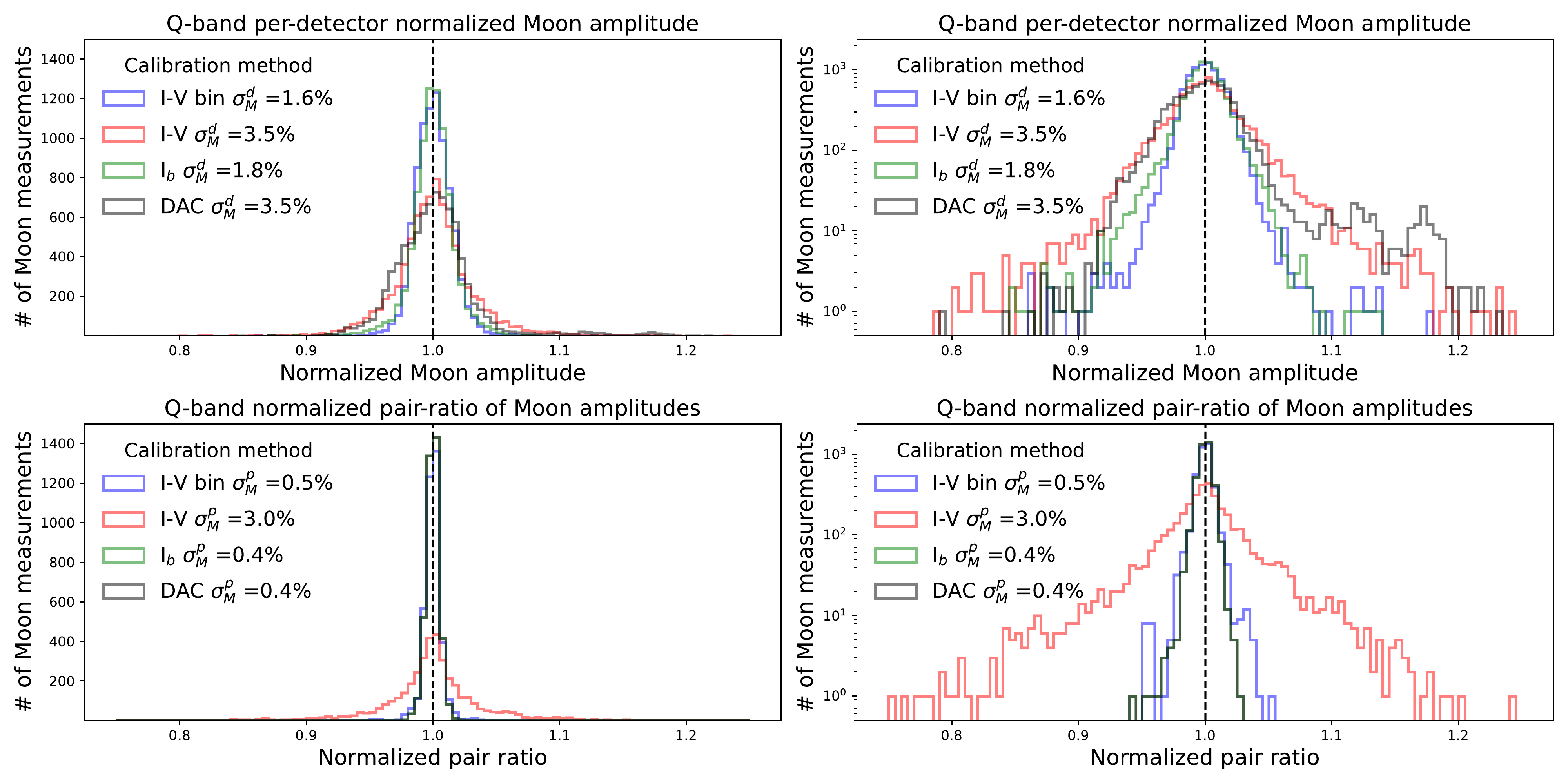}
\end{center}
\caption{Q-band Moon photometric calibration test histograms. \textbf{Top row:} The x-axis is the per-detector normalized Moon amplitudes across 208 Moon scans acquired between June 2016 and August 2020. The y-axis is the number of Moon measurements in a linear scale on the left and logarithmic scale on the right. Four Moon amplitude histograms are plotted, each corresponding to one of the four calibration models described in the text: \textcolor{blue}{$\square$}$I$-$V$~bin, \textcolor{red}{$\square$}$I$-$V$, \textcolor[rgb]{0.13,0.55,0.13}{$\square$}$I_{\mathrm{b}}$, and \textcolor{black}{$\square$}DAC. 
The $I_{\mathrm{b}}$ and $I$-$V$~bin methods yield the narrowest distribution and thus the most accurate calibrations.
These constraints are upper limits on the $I_{\mathrm{b}}$ and $I$-$V$~bin calibration accuracy, driven by uncertainty in the Moon antenna temperature model. 
\textbf{Bottom row:} By computing the normalized ratio of Moon amplitudes across detector pairs as shown in the bottom row, the uncertainty of the Moon antenna temperature model is eliminated, and we directly measure the calibration uncertainty between detector pairs. The x-axis is the normalized ratio of Moon amplitudes between detector pairs. The y-axis is the number of Moon ratios on a linear scale on the left and logarithmic scale on the right. The average ratios of each pair are normalized to one. The normalized ratios are then binned across the array, applying the four different calibration methods.  
The standard deviation values reported in the bottom row legends for ratios of detector pairs are divided by $\sqrt{2}$ for easier comparison to the standard deviations reported on the top row for single detectors.}
\label{fig:moon_calib}
\end{figure*}

The normalized Moon pair-ratio amplitudes shown in the bottom-row plots of Figure~\ref{fig:moon_calib} eliminate the uncertainty associated with the Moon antenna temperature model, since detector pairs measure the Moon at the same time (i.e., same Moon phase and angular diameter).
Additionally, the detectors in each pair share the same bias line and are therefore biased by the same $I_\mathrm{b}$ current, hence the ratios for the DAC and $I_\mathrm{b}$ calibration are identical.

The $I$-$V$~bin pair-ratio standard deviation indicates that the $I$-$V$~bin responsivities add uncorrelated error to the calibration at the $\sim $0.5\% level, comparable to the responsivity median standard error of 0.3\% calculated from the distribution of responsivities values in $I$-$V$ bins.

The small excess uncertainty may be attributed to: 
(1) additional error introduced by the beam-fitting algorithm used to extract the Moon amplitude. The Moon amplitude error of $\sim $0.2\% is based on the variance of the Moon map away from the peak signal and does not include systematic uncertainty from fitting the model beam profile to the Moon map; (2) systematic error in the $I$-$V$~bin responsivities not captured by the standard deviation in a bin; (3) error in the Moon model brightness temperature correction that persists in the pair ratio amplitudes due to systematic differences in the optical coupling of detector pairs.

\subsubsection{Variance of per-detector Moon amplitudes corrected by the Moon model uncertainty  $\sigma_\mathrm{M}^\mathrm{dM}$}
If we assume all $I$-$V$~bin calibration errors are stochastic, then the primary source of uncertainty for the $I$-$V$~bin and $I_\mathrm{b}$ photometric calibration tests is a ${\sqrt{(\sigma_\mathrm{M}^\mathrm{d} )^2-(\sigma_\mathrm{M}^\mathrm{p})^2}[I\mathrm{-}V~\mathrm{bin}]}= \sqrt{(1.6^2-0.5^2)} = 1.5\%$ error introduced by the Moon antenna temperature model.

Subtracting the Moon model uncertainty from the $I$-$V$ calibration yields a per-detector calibration uncertainty of ${\sigma_\mathrm{M}^\mathrm{dM}[I\mathrm{-}V] =3.1\%}$, consistent with the ${\sigma_\mathrm{M}^\mathrm{p}[I\mathrm{-}V] =3.0\%}$ per-detector calibration uncertainty estimated from the $I$-$V$ pair-ratio result.
The un-calibrated (DAC) per-detector variance corrected for the Moon model uncertainty ${\sigma_\mathrm{M}^\mathrm{dM}[\mathrm{DAC}]} =3.1\%$~(1-$\sigma$), indicates the level of detector gain variations across the selected Q-band Moon observations. 

The DAC distribution of Moon pair-ratios indicates a 1-$\sigma$ standard deviation of ${\sigma_\mathrm{M}^\mathrm{p}[\mathrm{DAC}]} =0.4\%$ in the relative variation of detector responsivity between  pairs, smaller than ${\sigma_\mathrm{M}^\mathrm{dM}[\mathrm{DAC}]} =3.1\%$. This implies the per-detector gain variations are dominated by common-mode responsivity changes across pairs.

\subsubsection{Variance of the detector array responsivities $\sigma^\mathrm{d}_\mathrm{S}$ and the relative gain calibration between detectors $\sigma^d_{\epsilon_j}$}
As expected, the per-detector DAC gain variations ($\sigma_\mathrm{M}^\mathrm{dM}$= 3.1\%) are similar to the variations in the per-detector responsivity values  $\sigma_\mathrm{S}^\mathrm{d}$[$I$-$V$~bin]$ = 3.3\%$.
This variance of per-detector responsivities  calculated for the subset of Moon observations considered in this photometric test is smaller than the $\sigma^\mathrm{d}_\mathrm{S} = 13\%$ reported in Table~\ref{tab:class_summary}. The high signal-to-noise Moon observation sample used for the photometric test selects the best observing conditions and excludes poor-quality data from periods where changes in optical loading would drive greater variation in the TES responsivities.
Also note that the per-detector gain variations across Moon observations are smaller than the differences in relative optical efficiency between detectors: $\sigma_\mathrm{\epsilon_j}^\mathrm{d} = 6.5\%(16.6\%)$, with $I$-$V$~bin calibration applied (uncalibrated).  
Differences in the relative optical efficiencies are due to detector characteristics that are stable across time, such as the detector bandpass and location on the focal plane. 

\begin{deluxetable}{ccccc}[ht]
\tablenum{4}
\tablehead{
\colhead{} & \colhead{\hspace{.35cm}$I$-$V$~bin}\hspace{.35cm} & \colhead{\hspace{.35cm}{$I$-$V$}}\hspace{.35cm} & \colhead{\hspace{.35cm}$I_{\mathrm{b}}$}\hspace{.35cm} & \colhead{\hspace{.35cm}DAC}\hspace{.35cm}
}
\startdata
$\sigma_\mathrm{M}^\mathrm{d}$ (\%) & 1.6  & 3.5  & 1.8  & 3.5  \\
$\sigma_\mathrm{M}^\mathrm{p}$ (\%) & 0.5  & 3.0 & 0.4  & 0.4  \\[4pt]
\hline
$\sigma_\mathrm{M}^\mathrm{dM}$ (\%) & 0.5  & 3.1  & 1.0  & 3.1  \\[4pt]
\hline
$\sigma^\mathrm{d}_\mathrm{S}$ (\%) & 3.3  & 4.4 & 3.0  & NA  \\
$\sigma^\mathrm{d}_{\epsilon_j}$ (\%) & 6.5  & 6.4  & 7.0  & 16.6  \\
\enddata
\caption{CLASS Q-band Moon photometric test results for the calibration methods: $I$-$V$~bin, $I$-$V$, $I_{\mathrm{b}}$, and DAC. Results are summarized in terms 1-$\sigma$ standard deviation of the measured distributions.
$\sigma_\mathrm{M}^\mathrm{d}$ is the standard deviation of the per-detector Moon amplitudes,
$\sigma_\mathrm{M}^\mathrm{p}$ is the standard deviation of the detector pair-ratio Moon amplitudes divided by $\sqrt{2}$,
$\sigma_\mathrm{M}^\mathrm{dM}$ is the standard deviation of the per-detector Moon amplitudes corrected by the Moon model uncertainty,
$\sigma^\mathrm{d}_\mathrm{S}$ is the standard deviation of the detector array responsivities,
and $\sigma^d_{\epsilon_j}$ is the standard deviation of the relative gain between detectors in the array.
The $\sigma_\mathrm{M}^\mathrm{p}$ and $\sigma_\mathrm{M}^\mathrm{dM}$ rows are similar for the $I$-$V$~bin, $I$-$V$, and $I_{\mathrm{b}}$ calibrations, because both rows remove the Moon model uncertainty---in $\sigma_\mathrm{M}^\mathrm{p}$ by taking the ratio across detector pairs, and in $\sigma_\mathrm{M}^\mathrm{dM}$ by subtracting in quadrature the estimated 1.5\% Moon uncertainty from $\sigma_\mathrm{M}^\mathrm{d}$.
}
\label{tab:class_moon_test}
\end{deluxetable}

\subsection{NEP calibration test}

By comparing the expected and measured noise amplitudes of detector data, we can test the accuracy of the detector calibration and identify possible systematic biases or excess noise sources.

The detector NEP is measured from TODs that are calibrated to power units through TES responsivity. The measured NEP values are compared to a robust TES NEP model based on measured detector bandpass and TES properties. 
Discrepancies between the average measured NEP and the NEP model can be attributed to excess noise sources or a responsivity calibration systematic error, particularly when the measured NEP falls below the model's prediction.
This comparison between modeled and measured NEP validates the $I$-$V$~bin responsivity calibration and indicates that the $I_{\mathrm{b}}$ responsivity is biased high.

The columns in Table~\ref{tab:class_nep_test} show the average NEP and $P_{\gamma}$ for the four CLASS frequency bands.
The NEP model average ($\overline{\mbox{NEP}}_\mathrm{m}$) is based on measured dark NEPs, Rayleigh\textendash{Jeans} center frequencies, and bandwidths reported in \citep{sumit_4year_det}.
$\overline{\mbox{NEP}}_{\mathrm{S}_\mathrm{I}}$ is the average NEP computed using the $I_\mathrm{b}$ responsivity calibration.
$\overline{\mbox{NEP}}_\mathrm{S}$ is the average NEP computed using the $I$-$V$~bin responsivity calibration.
For all four bands, $\overline{\mbox{NEP}}_{\mathrm{S}_\mathrm{I}}$ is lower than the theoretical lower limit set by $\overline{\mbox{NEP}}_\mathrm{m}$, indicating a bias in the calibration method.
$\overline{\mbox{NEP}}_\mathrm{S}$ is similar to or above $\overline{\mbox{NEP}}_\mathrm{m}$.
Excess NEP can be attributed to additional noise sources, such as instrument or atmospheric $1/f$ systematics. 

\begin{deluxetable}{ccccc}[ht]
\tablenum{5}
\label{tab:class_nep_test}
\tablehead{
\colhead{} & \colhead{\textbf{40 GHz}} & \colhead{\textbf{90 GHz}} & \colhead{\textbf{150 GHz}} & \colhead{\textbf{220 GHz}}
}
\startdata
$\overline{P}_{\gamma}$ ($\SI{}{\pico\watt}$) & 1.2  & 3.0  & 3.6  & 7.4  \\
$\overline{\mbox{NEP}}_\mathrm{m}$ ($\SI{}{\atto\watt\sqrt{\second}}$) & 15.8  & 29.8  & 34.2  & 53.2  \\
$\overline{\mbox{NEP}}_{\mathrm{S}_\mathrm{I}}$ ($\SI{}{\atto\watt\sqrt{\second}}$) & 15.0  & 25.0  & 30.0  & 47.0  \\
$\overline{\mbox{NEP}}_\mathrm{S}$ ($\SI{}{\atto\watt\sqrt{\second}}$) & 15.8  & 31.7  & 35.0  & 57.3  \\
\enddata
\caption{CLASS NEP responsivity test results including the average array optical loading and NEP for the four CLASS frequency bands. Only test data with the VPM turned off is used for this analysis. This allows us to focus on detector performance without having to account for modulation related systematics. $\overline{P}_{\gamma}$ is the array average optical loading. $\overline{\mbox{NEP}}_\mathrm{m}$ is the expected NEP based on the detector bandpass and dark NEP. $\overline{\mbox{NEP}}_{\mathrm{S}_\mathrm{I}}$ is measured NEP using $I_\mathrm{b}$ responsivities. 
$\overline{\mbox{NEP}}_\mathrm{S}$ is the measured NEP using $I$-$V$~bin responsivities.
Note that the NEPs from the $I_\mathrm{b}$ calibration are lower than expected values from the model, indicating that it biases low the absolute detector power calibration. 
}
\end{deluxetable}

\section{Conclusion}

We introduced a TES model operating at equilibrium in the low electrical-inductance limit that allowed us to directly relate the
detector responsivity calibration and optical time constant to the measured TES current $I$ and the applied bias
current $I_{\mathrm{b}}$. 
A novel TES bolometer calibration method based on binning $I$-$V$ data was applied to CLASS data acquired between June 2016 and August 2020, improving accuracy compared to single $I$-$V$ calibration.
This $I$-$V$~bin calibration method can be applied to existing or future CMB data sets where many $I$-$V$ measurements were acquired as part of the observing strategy.

We find that a simple TES calibration model based on the applied detector bias current yields precise per-detector TES power calibration factors for the CLASS Q-band array, but biases the absolute power calibration low. We conjecture that the observed high precision of the $I_{\mathrm{b}}$ per-detector calibration is a result of the uniformity of the Q-band detector parameters and the relatively stable atmospheric emission at \SI{40}{\giga\hertz}.

The atmospheric opacity model developed for the CLASS frequency bands reproduces the measured optical loading in each band as a function of PWV, corroborating the applied atmospheric gain correction model. We have described the calibration of CLASS TES bolometer data including factors normalizing for: 1) detector gain variations across time due to changes in optical loading and detector bias voltage; 2) optical efficiency variations across detectors; 3) changes in atmospheric opacity; and 4) an absolute temperature calibration based on Moon and planet observations.

The accuracy of the calibration pipeline is tested using Q-band Moon observations as a photometric standard and by comparing the measured detector noise versus NEP models at all four CLASS frequency bands.

The $I$-$V$~bin calibration method yields a median per-detector TOD gain uncertainty of 0.3\% in Q-band data, which is corroborated by using high signal-to-noise Moon observations as a photometric standard. The CLASS CMB data set is composed of thousands of day-long detector TODs; hence, we expect the demonstrated calibration uncertainty to contribute a negligible systematic error to the resulting maps and power spectra to be discussed in upcoming CLASS publications.
\begin{acknowledgments}
We acknowledge the National Science Foundation Division of Astronomical Sciences for their support of CLASS under Grant Numbers 0959349, 1429236, 1636634, 1654494, 2034400, and 2109311. We thank Johns Hopkins University President R. Daniels and the Deans of the Kreiger School of Arts and Sciences for their steadfast support of CLASS. We further acknowledge the very generous support of Jim and Heather Murren (JHU A\&S ’88), Matthew Polk (JHU A\&S Physics BS ’71), David Nicholson, and Michael Bloomberg (JHU Engineering ’64). The CLASS project employs detector technology developed in collaboration between JHU and Goddard Space Flight Center under several previous and ongoing NASA grants. Detector development work at JHU was funded by NASA cooperative agreement 80NSSC19M0005. CLASS is located in the Parque Astron\'omico Atacama in northern Chile under the auspices of the Agencia Nacional de Investigaci\'on y Desarrollo (ANID). We acknowledge scientific and engineering contributions from Max Abitbol, Fletcher Boone, Jay Chervenak, Lance Corbett, David Carcamo, Mauricio D\'iaz, Ted Grunberg, Saianeesh Haridas, Connor Henley, Ben Keller, Lindsay Lowry, Nick Mehrle, Grace Mumby, Diva Parekh, Isu Ravi, Daniel Swartz, Bingjie Wang, Qinan Wang, Emily Wagner, Tiffany Wei, Zi\'ang Yan, Lingzhen Zeng, and Zhuo Zhang. For essential logistical support, we thank Jill Hanson, William Deysher, Miguel Angel D\'iaz, Mar\'ia Jos\'e Amaral, and Chantal Boisvert. We acknowledge productive collaboration with Dean Carpenter and the JHU Physical Sciences Machine Shop team. I.L.P. gratefully acknowledges support from the Horizon Postdoctoral Fellowship.
S.D. is supported by an appointment to the NASA Postdoctoral Program at the NASA Goddard Space Flight Center, administered by Oak Ridge Associated Universities under contract with NASA. S.D. acknowledges support under NASA-JHU Cooperative Agreement 80NSSC19M005. 
R.R. acknowledges partial support from CATA, BASAL grant AFB-170002, and 
CONICYT-FONDECYT through grant 1181620. Z.X. is supported by the Gordon and Betty Moore Foundation through grant GBMF5215 to the Massachusetts Institute of Technology.
\end{acknowledgments}

\software{\texttt{NumPy} \citep{numpy}, \texttt{SciPy} \citep{scipy}, \texttt{Matplotlib} \citep{matplotlib}}

\appendix
\section{Electrothermal speed-up factor}
\label{appendixA}

\begin{figure}
\tikzstyle{block} = [draw, fill=blue!10, rectangle,
    minimum height=3em, minimum width=6em]
\tikzstyle{sum} = [draw, fill=blue!10, circle, node distance=1cm]
\tikzstyle{input} = [coordinate]
\tikzstyle{output} = [coordinate, node distance=3.5cm]
\tikzstyle{pinstyle} = [pin edge={to-,thin,black}]
\begin{tikzpicture}[auto, node distance=2cm,>=latex']
    \node [input, name=input] {};
    \node [sum, right of=input] (sum) {};
    \node [block, right of=sum, node distance = 2.7cm] (controller) {$A$};
    \node [output, right of=controller] (output) {};
    \node [block, below of=controller] (measurements) {$B$};
    \draw [draw,->] (input) -- node {$X$} (sum);
    \draw [->] (sum) -- node 
    {$E$} (controller);
    \draw [->] (controller) -- node [name=y] {$Y$}(output);
    \draw [->] (y) |- (measurements);
    \draw [->] (measurements) -| node[pos=0.99] {$+$}
        node [near end] {} (sum);
\end{tikzpicture}
\caption{Feedback block diagram describing the variables discussed in Appendix~\ref{appendixA}}
\label{fig:simple_block}
\end{figure}
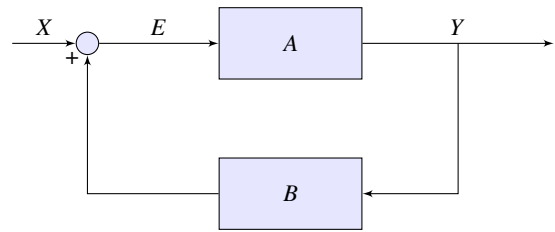

This appendix elaborates on the TES time constant speed-up factor attributed to the TES electrothermal feedback. 
It describes how to arrive at Equation~\ref{eqn:tau_factor} starting from the feedback circuit shown in Figure~\ref{fig:simple_block}.

Consider the ratio $Z$ of the output $Y$ to input $X$ of the block diagram in Figure~\ref{fig:simple_block}:
\begin{equation}
Z = \frac{Y}{X}.
\label{eqn:ratioT}
\end{equation}
The block diagram implies that $X$ and $Y$ satisfy:
\begin{equation}
Y = A E, \mathrm{~and~} E = X+BY.
\label{eqn:yae}
\end{equation}
Combining equations~\ref{eqn:ratioT} and \ref{eqn:yae} to solve for $Z$, we arrive at
\begin{equation}
Z = \frac{AE}{E-BY} = \frac{A}{1-BY/E}. = \frac{A}{1-BA} = \frac{1}{1/A-B}.
\label{eqn:Z_AB}
\end{equation}
Now consider $Z$ to be the ratio of the output detector current $I$ to the input power $P_{\gamma}$ and assuming a single-pole detector temporal response  with effective time constant $\tau_{\gamma}$, then
\begin{equation}
Z = \frac{Z_o}{1+i\omega\tau_{\gamma}}.
\label{eqn:Z_tau}
\end{equation}
If the feedback loop is broken by removing $B$, then the open loop response of $A$ is a single-pole transfer function set by the TES thermal time constant $\tau = C/G$, where $C$ is the TES heat capacity and $G$ its thermal conductivity, therefore
\begin{equation}
A = \frac{A_o}{1+i\omega\tau}.
\label{eqn:A_tau}
\end{equation}
Typically, $\tau_{\mathrm{el}}< L/R \ll \tau$ is a good assumption, where $L/R$ is an upper limit ($\beta\sim0$ and $R_{\mathrm{sh}}/R\sim0$) on the TES electrical time constant $\tau_{\mathrm{el}}$ (see Equation~\ref{eqn_tel}). $L$ is the TES circuit inductance and $R$ the TES resistance.
In this low electrical-inductance limit, the electrothermal feedback has infinite bandwidth, and the transfer function of $B$ is a constant factor $B_o$.
Substituting equations~\ref{eqn:Z_tau} and \ref{eqn:A_tau} into \ref{eqn:Z_AB}, we obtain
\begin{equation}
\frac{Z_o}{1+i\omega\tau_{\gamma}} = \frac{A_o}{1+i\omega\tau-B_o A_o} = \frac{A_o/(1-B_o A_o)}{1+i\omega(\tau/(1-B_o A_o))}.
\label{eqn:Zo_Ao_tauo}
\end{equation}
From Equation~\ref{eqn:Zo_Ao_tauo}, we identify:
\begin{equation}
Z_o = \frac{A_o}{1-B_o A_o} \mathrm{~and~} \tau_{\gamma} =\frac{\tau}{1-B_o A_o},
\label{eqn:Zo_tauo}
\end{equation}
hence arriving at Equation~\ref{eqn:tau_factor}:
\begin{equation}\label{eqn:AoBo}
\frac{\tau_{\gamma}}{\tau} = \frac{1}{1- A_o B_o}.
\vspace{1mm}
\end{equation}
Equation~\ref{eqn:tau_factor} can be expressed in terms of the ideal TES DC responsivity $S^*$ and responsivity $S$ by identifying $A_o$ and $B_o$ in the block diagram of Figure~\ref{fig:tes_bol} as
\begin{equation}
A_o \equiv \left.\frac{\partial I}{\partial P_{\upvarphi}}\right|_{I_{\mathrm{b}}} = \Big(\left.\frac{\partial P_{\gamma}}{\partial I}\right|_{I_{\mathrm{b}}}+\left.\frac{\partial P_{\mathrm{J}}}{\partial I}\right|_{I_{\mathrm{b}}}\Big)^{-1}= \Big(\frac{1}{S}-\frac{1}{S^*}\Big)^{-1},
\label{eqn:Ao_S}
\end{equation}
and
\begin{equation}
B_o \equiv\left.\frac{\partial P_{\mathrm{J}}}{\partial I}\right|_{I_{\mathrm{b}}}= R_{\mathrm{sh}}(I_{\mathrm{b}}-2 I)= -\frac{1}{S^*}.
\label{eqn:Bo_S}
\end{equation}
Combining equations~\ref{eqn:AoBo}, \ref{eqn:Ao_S}, and \ref{eqn:Bo_S} yields
\begin{equation}
\frac{\tau_{\gamma}}{\tau} = \frac{1/A_o}{1/A_o-B_o} = \frac{1/S-1/S^*}{1/S-1/S^*+1/S^*} = 1-\frac{S}{S^*}.
\label{eqn:tau_S}
\end{equation}
Equating the time constant ratio to the responsivity values allows us to quickly compute the electrothermal speed-up factor with the responsivity estimates already extracted from every $I$-$V$ measurement. 

\bibliography{ref}{}
\bibliographystyle{aasjournal}
\end{document}